\documentclass{aa}  

\usepackage{amsmath}
\usepackage{graphicx}
\usepackage{txfonts}
\usepackage{natbib}
\usepackage{soul}

\begin{document}

   \title{What happened before?}

   \subtitle{The disks around the precursors of young Herbig Ae/Be stars.}

   \author{P.-G. Valeg\r{a}rd\inst{1}
          \,
          L.B.F.M. Waters\inst{2,3}
          \,
          C. Dominik\inst{1}
          }

   \institute{Anton Pannekoek Institute for Astronomy (API), University of Amsterdam, Science park 904, 1098 XH Amsterdam, The Netherlands\\
              \email{p.g.valegard@uva.nl}
         \and
         Institute for Mathematics, Astrophysics \& Particle Physics, Radboud University, P.O. Box 9010, MC 62, NL-6500 GL Nijmegen, the Netherlands
         \and
         SRON, Sorbonnelaan 2, 3484CA Utrecht, The Nederlands
             }

   \date{Received October 30, 2020; accepted April 15, 2021}

 
  \abstract
   {Planets form in circumstellar disks around pre-main sequence stars. A key question is how the formation and evolution of protoplanetary disks depends on stellar mass. Studies of circumstellar disks at infrared and submm wavelengths around intermediate mass Herbig Ae/Be stars have revealed disks structures such as cavities, gaps and spiral arms. The Herbig Ae/Be stars represent an older population of intermediate mass pre-main sequence stars. Since intermediate mass pre-main sequence stars evolve towards the main sequence on timescales comparable to typical disk dissipation timescales, a full picture of disk dispersal in intermediate mass pre-main sequence stars must include the intermediate mass T-Tauri stars.} 
   {We seek to find the precursors of the Herbig Ae/Be stars in the solar vicinity within 500 pc from the Sun. We do this by creating an optically selected sample of intermediate mass T-Tauri stars (IMTT stars) here defined as stars of masses $1.5 M_{\odot}\leq M_* \leq 5 M_{\odot}$ and spectral type between F and K3, from literature.}
   {We use literature optical photometry (0.4-1.25$\mu$m) and distances determined from \textit{Gaia} DR2 parallax measurements together with Kurucz stellar model spectra to place the stars in a HR-diagram. With Siess evolutionary tracks we identify intermediate mass T-Tauri stars from literature and derive masses and ages. We use Spitzer spectra to classify the disks around the stars into Meeus Group I and Group II disks based on their [F$_{30}$/F$_{13.5}$] spectral index. We also examine the 10$\mu$m silicate dust grain emission and identify emission from Polycyclic Aromatic Hydrocarbons (PAH). From this we build a qualitative picture of the disks around the intermediate mass T-Tauri stars and compare this with available spatially resolved images at infrared and at sub-millimeter wavelengths to confirm our classification. }
   {We find 49 intermediate mass T-Tauri stars with infrared excess. The identified disks are similar to the older Herbig Ae/Be stars in disk geometries and silicate dust grain population. The detection frequency of PAHs is higher than from disks around lower mass T-Tauri stars but less frequent than from Herbig Ae/Be disks. Spatially resolved images at infra-red and sub-mm wavelengths suggest gaps and spirals are also present around the younger precursors to the Herbig Ae/Be stars.}
   {Comparing the timescale of stellar evolution towards the main sequence and current models of protoplanetary disk evolution the similarity between Herbig Ae/Be stars and the intermediate mass T-Tauri stars points towards an evolution of Group I and Group II disks that are disconnected, and that they represent two different evolutionary paths.}


   \maketitle
%
\section{Introduction}
    Planets are formed in circumstellar disks around young pre-main sequence stars (PMS stars) or perhaps, even in the earlier proto-stellar phase \citep{2020Natur.586..228S}. Angular momentum conservation during the collapse of the molecular cloud around the proto-stellar core forces material to form a disk around the forming star. The disk material dissipates over a few Myrs as the material is accreted onto the star, lost into space by disk winds and/or jets, and accreted into planets. The circumstellar disk evolves from a gas rich disk into a transitional disk and eventually a gas poor debris disk.
    This general scenario has gained substantial observational support over the last 10 years, as more spatially resolved observations around PMS stars have become available. Observations at mm and sub-mm wavelengths with instruments such as for example ALMA, SPHERE, GPI and SUBARU have revealed circumstellar disks with spiral arms, central cavities, concentric circular gaps and warps in the disk indicative of interactions with planetary bodies \citep{2020arXiv200105007A,2017AA...603A..21G,2018ApJ...869L..41A,2018ApJ...869L..42H,2018ApJ...869L..43H, 2020AA...633A..82G,2018AA...619A.171B, 2019ApJ...872..122M}. Other explanations for these features have also been suggested such as gravitational instabilities, snow lines and binary interactions \citep{ 2020ARA&A..58..483A, 2019ApJ...872..112V}.

    A key questions is to understand how the formation and evolution of planetary systems depends on stellar mass. Can we connect the properties of planet-forming disks to those of mature planetary systems, and what role does the stellar mass play? When do the features seen in observations of planetary disks, spirals, gaps and cavities, form and what do they say about the planetary formation process?

    Two main types of PMS stars have traditionally been observed to study the circumstellar disks and their evolution. The T-Tauri stars (spectral class F and later) \citep{1945ApJ...102..168J,1962AdAA...1...47H,1983AA...126..438B} and the Herbig Ae/Be stars (hereafter HAeBe, spectral class B and A) \citep{1960ApJS....4..337H,1983AA...126..438B}. Overlapping in mass, the T Tauri stars with masses $M \geq 1.5 M_{\odot}$ will evolve through the Hertzsprung-Russel diagram (HR-diagram) to become HAeBe stars. The studies of circumstellar disks using HAeBe stars only, however, have introduced a bias in our view of planet formation and disk dissipation in this mass range \citep[see also][]{2018AA...620A.128V}. 
    
    In order to have a full observational view of disk evolution and planet formation for intermediate mass stars, samples must be constructed that contain both HAeBe stars and their precursors. Several papers have mentioned these T Tauri stars as a separate group. \citet{1988cels.book.....H} used a special category called 'su' T Tau stars, defined as stars with a spectral appearance similar to the T Tauri star SU Aur. \citet{1994AJ....108.1906H} used the term Early Type T Tauri stars or ETTS, containing both HAeBe stars and T Tauri stars. To our knowledge \citet{2004AJ....128.1294C} was the first paper dedicated to and naming the intermediate mass T Tauri stars (hereafter IMTT stars). \citet{2017AA...608A..77L} studied magnetic fields in IMTT stars and defined the class as PMS stars with masses $ 1 M_{\odot} \leq M \leq 4 M_{\odot}$. These papers contain a relatively small number of objects and uses different mass ranges. The IMTT stars have so far been discussed as a part of a much larger samples of T Tauri stars. For the HAeBe stars, samples exists and are widely studied (e.g. Herbig 1960; The et al. 1994; Acke {\&} van den Ancker 2006; Juh\'{a}sz et al. 2010; Garufi et al. 2017; Varga et al 2018a; Vioque et al 2018).

    It is therefore useful to construct a sample of IMTT stars from existing literature and study their properties, so that a more complete picture of planetary formation and disk evolution for intermediate mass PMS stars can be constructed and compared to the lower mass stars.

    In this paper we have identified IMTT stars by searching the literature for luminous T Tauri stars in the solar neighborhood that may be of intermediate mass, defined by their mass $1.5 M_{\odot} \leq M \leq 5 M_{\odot}$ and spectral type later than F. We construct spectral energy distributions (SEDs) and use literature values of spectral types and effective temperatures, together with \textit{Gaia} DR2 data in order to place the stars into the HR diagram. We discuss the properties of the stars and their disks in the context of what is known for the HAeBe stars by examining disk geometry, silicate dust grains and Polycyclic Aromatic Hydrocarbons (PAH).
    \citet{2001AA...365..476M} used the shape of the infrared and millimeter spectrum to classify the protoplanetary disks around HAeBe stars into two groups. In Group I (transitional disks) and Group II based on their SED. Since then many papers have used this classification and noted differences between the two groups. In this paper we will also use the Meeus classification so that a meaningful comparison with the HAeBe literature can be made.

    This paper is organized as follows: In Section 2 we discuss the construction and refining of the sample. In section 3 we describe our analysis of the disk properties, focusing on spatially unresolved data. Section 4 discusses the properties of IMTT disks in the context of disk evolution and dissipation in intermediate mass PMS stars, and in Section 5 we summarize the main conclusions of our study.

\section{Sample selection}

\subsection{Approach}
    The optically selected sample has been constructed by searching literature for IMTT stars in the solar neighbourhood. The selection procedure takes place in 4 steps. First an initial search for T-Tauri stars in the SIMBAD database. Then in a second step we remove stars that lie beyond a chosen distance maximum. We use a simple method to obtain a first estimate of luminosity in the third step in order to remove most of the low mass T-Tauri stars ($<1.5 M_{\odot}$). Finally, in a forth step, the luminosity of the remaining stars are then determined more precisely by fitting a Kurucz stellar model to the photometry, allowing an accurate placement in the HR-diagram.
    
\subsection{Initial selection of stars from literature}    
    We searched the SIMBAD database for T-Tauri stars (keyword TT*) with spectral types between $F0-K3$. Earlier type stars one would contaminate the sample by including the Herbig stars; with later type stars than K3 estimating the mass correctly is challenging with our methodology because of the proximity of the pre-main sequence evolutionary tracks which run close together and are almost vertical in the particular luminosity range we are interested in. The search turned up 623 stars. We have also included HAeBe stars of spectral type F and G from \citet{2018AA...620A.128V} (2 stars) as well as the IMTT stars in the work by \citet{2019AA...622A..72V} (14 stars) and \citet{2004AJ....128.1294C} (2 stars).
    
\subsection{Distance determination and distance limit}    
    The disks around the stars should be observable in scattered light in the hope that disk features such as gaps, spirals and shadows in the disk can be observed. Gaps in the disk typically have an average width of $ \leq 10 AU$ \citep{2018ApJ...869L..41A}. With the current resolution of scattered light instruments such as VLT-SPHERE, that have a resolution limit of 22-27 mas \citep{2019AA...631A.155B} these features would not be easily detected at distances beyond 450 pc. Since disks themselves can be detected in scattered light at greater distances and the number of high mass star-forming regions in the vicinity of the sun is low we decided to include the Orion star forming region. Therefore we extend the distance limit to 500 pc, keeping in mind that features in the most distant disks will be difficult to resolve. 
    In the second step, we obtained the distances to each star by bayesian inference from \textit{Gaia DR2} parallaxes by \citet{2018AJ....156...58B}. This is an important asset for our method determining stellar parameters because of the high precision in the parallaxes measured by \textit{Gaia}. The distance uncertainties by Bayesian inference presented in \citet{2018AJ....156...58B} for the final sample of stars are typically between 2-4\%. The uncertainties in luminosity for the stars is therefore dominated by the uncertainty in the determined $T_{\rm eff}$. If no distance was available we used Hipparcos parallaxes. For stars where none of these options where available we searched literature for a distance measurement or estimate using associations of stars. Stars where no distance reference could be found or that where further away then 500 pc where then removed from the sample. We have used all Gaia parallaxes in our selection steps, i.e. also those with RUWE above 1.4. We will come back to this point in section 2.5 when discussing the final sample.
 
\subsection{Removal of lower mass T-Tauri stars}
    The third step in the selection process aims to remove the bulk of the low mass T-Tauri stars. We first look for an accurate spectral class classification in literature or an accurate determination of $T_{\rm eff}$. Stars that have a classification as FU Ori stars were immediately removed because of their high variability from outbursts and because the flux at optical wavelengths are dominated by their accretion disks. 
    
    Pre-main sequence evolutionary tracks partially overlap with post-main sequence tracks. Stars identified as post-main sequence in the literature in this step were therefore removed from the sample. The remaining number of post-main sequence stars in our sample should then be very small. Infrared excess is not expected for single intermediate mass stars that have evolved to sub-giants, as these stars do not yet have dusty stellar winds. 
    
    We then collected optical B and V-band photometry using catalogs with good coverage. These catalogs included: NOMAD \citep{2005yCat.1297....0Z}, UCAC4 \citep{2012yCat.1322....0Z}, APASS DR2 \citep{2016yCat.2336....0H}, UBVRIJKLMNH Photoelectric catalog \citep{1978AAS...34..477M}, Tycho2 catalog \citep{2000yCat.1259....0H}, EPIC \citep{2017yCat.4034....0H} and \citet{2006ApJ...653..657M}.  Using the table of standard stars from \citet{2013ApJS..208....9P} we obtain the intrinsic (B-V) for each spectral class and temperature as well as the bolometric corrections (BC) in V-band, BC$_{V}$. Assuming an $R_{v}$=3.1 we estimated the extinction, $A_{v}$ and the luminosity from the V-band magnitude using the distance modulus. We then set a lower luminosity limit of 2.1 $L_{\odot}$ based on the lowest luminosity of a \citet{2000AA...358..593S} stellar evolutionary track for a star of 1.5 $M_{\odot}$ with solar type metallicity (Z=0.01). Stars that are likely to be low mass T-Tauri stars are in this way removed. This cut the sample to 122 candidates.


   
   
   \begin{figure*}[ht]
        \includegraphics[width=\textwidth]{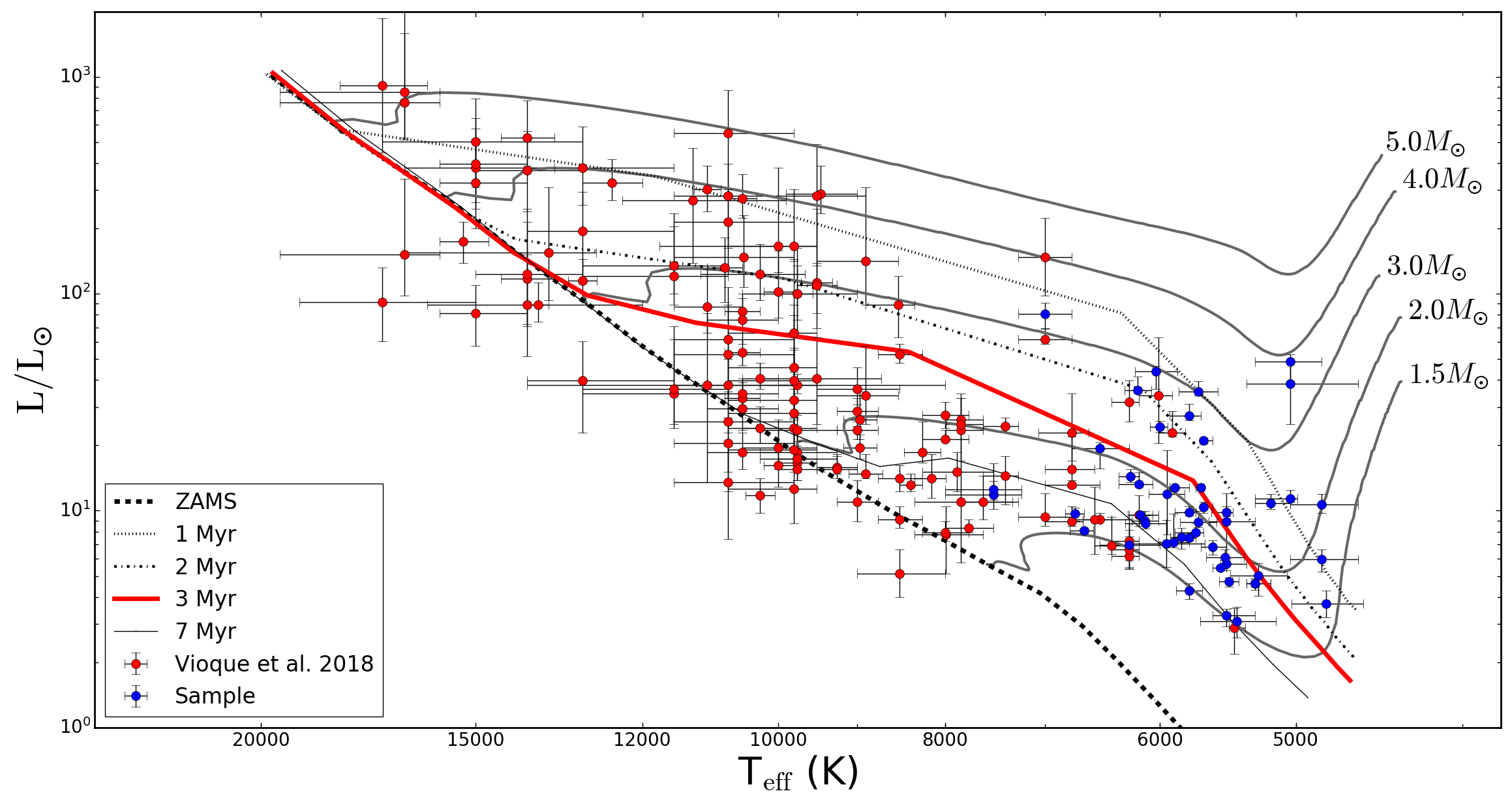}
        \caption{This HR-diagram shows the IMTT stars sample (blue) and the HAeBe sample from Vioque et al. 2018 (red) that lies within the same mass range as the IMTT stars. The pre-main sequence evolutionary tracks (grey-solid) are from \citet{2000AA...358..593S}  The isochrones shown are for 1 Myr (dotted), 2 Myr (dash-dotted), 3 Myr (red-solid) and for 7 Myr (black-solid) and the ZAMS (thick-dashed) is defined by the location when the nuclear luminosity provides 99$\%$ of the total stellar luminosity.}
        \label{fig:C-HRdiagram}
    \end{figure*}

\subsection{Final selection}
    
    The fourth and final step in our source selection procedure involves placing the stars accurately in the HR diagram by means of SED fitting. In order to do so the extinction and luminosity needs to be accurately determined. Together with $T_{\rm eff}$ the mass can be determined from evolutionary tracks in the HR diagram. Photometry used in this process covered 0.4-1.25 $\mu$m to avoid excess from accretion in the UV and infrared excess from circumstellar material. 
    
    We used \textit{Gaia} DR2 G-band photometry to examined the relative error in magnitude from the \textit{Gaia} database \citep{2018AA...616A...1G} and conclude that most stars have low to moderate variability (at least during the \textit{Gaia} measurement period) of typically around 1\% with a few exceptions up to around 6\%. Some variability has to be expected because of the presence of possibly occulting circumstellar material. Therefore, we use the maximum measured brightness at each wavelength.

    A Kurucz model spectrum was fitted to the optical photometry using the iterative method of \citet{2016AA...586A.103W} (see Appendix A in said paper). The method requires the following starting input parameters: $M_*$, distance , $T_{\rm eff}$, $L_*$ , $R_v$ and $A_v$. We use (as derived in previous subsection) the estimates for $A_v$, $L_*$ and the literature values for $T_{\rm eff}$.  The initial $M_*$ was set to $2M_{\odot}$ and the $R_v=3.1$ for all stars (except 3 highly reddened sources, EM*SR $21$, Haro $1-6$ and LkH${\alpha}$ 310, were $R_v = 5$ in the end gave a better stellar photospheric fit). The routine starts with calculating a $\log(g)$ assuming said starting parameters. This $\log(g)$ value is used to choose a Kurucz stellar spectrum. The spectrum is then reddened and fitted to the observed photometry, convolving the model spectrum with the filter transmission curves for each of the photometric bands to produce the integrated model flux in each band \citep{2019PASP..131f4301W}. The resulting $L_*$ and $A_v$ is then fed back into the program, keeping $T_{\rm eff}$ fixed. The stellar mass is refined for each iteration by deriving the value using \citet{2000AA...358..593S} pre-main sequence stellar evolution models. This process leads to a more accurate value $\log(g)$ and therefore a better model spectrum selection for the following iteration. The routine converges very quickly (typically in 2-3 iterations) to determine $M_*$, $L_*$ and $E(B-V)$. \footnote{We found that for HD 34700 the iteration finished with a negative extinction, it was kept in the sample by adjusting the spectral type by a half subclass \citep[i.e. we re-run the fit using a spectral type of F9 from][]{2013ApJS..208....9P} }  We run this procedure not only for the $T_{\rm eff}$ value, but also for the $T_{\rm eff}$ range allowed by the uncertainty, in order to derive the corresponding range in luminosity.

    Stars that were found to be below 1.5 $M_{\odot}$ or were no fit could be made with the available photometry were removed. We also remove stars that have no infrared excess, i.e. the photometry follows the expected stellar photospheric emission. We do not apply a strict upper bound for the luminosity as a selection criterion.  

    Some of the \textit{Gaia} parallaxes in our final sample are less reliable. We consider the parallaxes with re-normalized unit weight error, RUWE$\leq$ 1.4 as 'good'\footnote{see technical note GAIA-C3-TN-LU-LL-124-01, in the \textit{Gaia DR2} release documentation.}. As a final step in our procedure, we therefore carefully check the distances for 10 stars in our final sample for which the \textit{Gaia} RUWE was found to be $> 1.4$. In appendix A a comparison is presented between the \textit{Gaia} distance of these 10 stars and other distance determinations in the literature. Appendix A also presents our final value adopted for the distances of these stars.

    A few notes on specific targets are on order:
    \begin{itemize}
        \item Two stars in the sample are classified as late type A-stars (HD 142666 and HD 144432). They have remained in the sample because their uncertainty in $T_{\rm eff}$ overlaps with our sample criteria.
        \item
        We note that the 70 $\mu$m and sub-mm fluxes measured for HBC 502 \citep{2008AJ....135..966F,2001ApJ...556..215M} are very high. \citet{2010ApJ...708.1107M} suggest that this may be due a spatial coincidence with a class 0/I object. However, the optical and IR photometry is dominated by the IMTT star and therefore we kept this object in our list.
        \item
        For UX Tau we were unable to fit a model spectra. We adopted stellar parameters from literature \citep{2019ApJ...872..158A,2017AA...603A..74C,2010ApJ...717..441E,2009ApJ...704..531K}. Using these values we have kept it in the analysis because of its position in the HR-diagram.
        \item
         We also remove BP Psc, which is discussed to maybe be a T-Tau star ($\sim 80$ pc) or a post-MS giant ($\sim 300$ pc) \citep{2017MNRAS.466L...7D} . The distance from \citet{2018AJ....156...58B} release puts it at $\sim$360 pc suggesting a post main-sequence nature.
    \end{itemize}

\subsection{Final sample}

    We find 49 sources matching our selection criteria for IMTT stars, that also show evidence of infrared excess. An overview of the sample can be seen in table \ref{tbl:Star_Overview} together with the literature reference for their spectral class. The HR-diagram for the full sample is presented in figure \ref{fig:C-HRdiagram}. The majority of the IMTT stars found are located in the Orion star forming complex. Other regions include Cepheus, Chameleon, Lupus, Ophiucus, Perseus and Taurus-Auriga star forming region. The richness of sources in Orion may be due to a high abundance of these IMTT stars in this type of starfoming region also other star forming regions have ages that imply that intermediate mass PMS have already lost their disks. 10 of the stars have previously been described as T-Tau stars while the remaining have previously been identified as intermediate mass stars (IMTT or HAeBe stars) (see table \ref{tbl:Star_TTref}). We present the stellar parameters for these stars as found by the method described in the previous subsections in table \ref{tbl:Stellarparam}.

\begin{table*}
\caption{{The final IMTT star sample. References are to spectral class and/or $T_{\rm eff}$, the stars were $T_{\rm eff}$ is marked with an 'a' are converted from their literature spectral type to temperature using \citet{2013ApJS..208....9P}. }}
\label{tbl:Star_Overview}
\small
\begin{tabular}{llrrrrrrlll}
\\
\hline \hline
\multicolumn{2}{c}{Catalogue name} & \multicolumn{3}{c}{Right Ascension} & \multicolumn{3}{c}{Declination} &  SpT & T\textsubscript{eff} & Ref. to\\
Name & 2MASS & hh & mm & ss & deg & amin & asec & & [K] & SpT or $T_{\rm eff}$\\
\hline 
BX Ari       & J02581122+2030037 & 2  & 58 & 11.23 & 20  & 30  & 3.15  & K2      & 5040\tablefootmark{a} & \citet{2014ApJ...786...97H} \\
HBC 338      & J03254982+3110237 & 3  & 25 & 49.83 & 31  & 10  & 23.84 & G8      & 5490\tablefootmark{a} & \citet{2003AA...402..963M}  \\
LkH$\alpha$ 330     & J03454828+3224118 & 3  & 45 & 48.28 & 32  & 24  & 11.85 & F7      & 6240\tablefootmark{a} & \citet{2014ApJ...786...97H} \\
RY Tau       & J04215740+2826355 & 4  & 21 & 57.41 & 28  & 26  & 35.53 &         & 5945       & \citet{2004AJ....128.1294C} \\
T Tau        & J04215943+1932063 & 4  & 21 & 59.43 & 19  & 32  & 6.44  &         & 5700       & \citet{2004AJ....128.1294C} \\
UX Tau A     & J04300399+1813493 & 4  & 30 & 4     & 18  & 13  & 49.44 & G8      & 5490\tablefootmark{a} & \citet{2019ApJ...872..158A} \\
HQ Tau       & J04354733+2250216 & 4  & 35 & 47.33 & 22  & 50  & 21.64 & K0      & 5280\tablefootmark{a} & \citet{2012ApJ...745..119N} \\
HBC 415      & J04355415+2254134 & 4  & 35 & 54.16 & 22  & 54  & 13.4  & G2      & 5770\tablefootmark{a} & \citet{2014ApJ...786...97H} \\
SU Aur       & J04555938+3034015 & 4  & 55 & 59.39 & 30  & 34  & 1.5   & G4      & 5680\tablefootmark{a} & \citet{2014ApJ...786...97H} \\
HD 34700     & J05194140+0538428 & 5  & 19 & 41.41 & 5   & 38  & 42.78 & G0 Ive  & 5920\tablefootmark{a} & \citet{2001AA...378..116M}  \\
CO Ori       & J05273833+1125389 & 5  & 27 & 38.34 & 11  & 25  & 38.92 &         & 6030       & \citet{2004AJ....128.1294C} \\
HD 35929     & J05274279$-$0819386 & 5  & 27 & 42.79 & -8  & 19  & 38.45 &         & 7000       & \citet{2015MNRAS.453..976F} \\
PDS 115      & J05281785+0110061 & 5  & 28 & 17.85 & 1   & 10  & 6.12  & G2      & 5770\tablefootmark{a} & \citet{2008MNRAS.387.1335R} \\
GW Ori       & J05290838+1152126 & 5  & 29 & 8.39  & 11  & 52  & 12.65 &         & 5700       & \citet{2019AA...622A..72V}  \\
V1650 Ori    & J05291144$-$0608054 & 5  & 29 & 11.44 & -6  & 8   & 5.4   &         & 6160       & \citet{2019AA...622A..72V}  \\
GX Ori       & J05300203+1213357 & 5  & 30 & 2.04  & 12  & 13  & 35.86 &         & 5410       & \citet{2004AJ....128.1294C} \\
RY Ori       & J05320993$-$0249467 & 5  & 32 & 9.94  & -2  & 49  & 46.77 &         & 6120       & \citet{2019AA...622A..72V}  \\
HBC 442      & J05341416$-$0536542 & 5  & 34 & 14.16 & -5  & 36  & 54.19 &         & 6170       & \citet{2006ApJ...653..657M} \\
SW Ori       & J05341574$-$0636046 & 5  & 34 & 15.75 & -6  & 36  & 4.68  & G8      & 5490\tablefootmark{a} & \citet{2013ApJ...764..114H} \\
V1044 Ori    & J05341646$-$0536455 & 5  & 34 & 16.46 & -5  & 36  & 45.64 &         & 5500       & \citet{2019AA...622A..72V}  \\
EZ Ori       & J05341856$-$0504479 & 5  & 34 & 18.57 & -5  & 4   & 47.77 &         & 5830       & \citet{2004AJ....128.1294C} \\
Brun 252     & J05342495$-$0522055 & 5  & 34 & 24.96 & -5  & 22  & 5.53  &         & 5890       & \citet{2019AA...622A..72V}  \\
V2149 Ori    & J05350519$-$0514503 & 5  & 35 & 5.21  & -5  & 14  & 50.37 &         & 6180       & \citet{2019AA...622A..72V}  \\
Brun 555     & J05351511$-$0444429 & 5  & 35 & 15.13 & -4  & 44  & 42.96 & K2      & 5040\tablefootmark{a} & \citet{1979ApJS...41..743C} \\
Brun 656     & J05352131$-$0512126 & 5  & 35 & 21.32 & -5  & 12  & 12.61 & G2 III  & 5770\tablefootmark{a} & \citet{1969ApJ...155..447W} \\
V815 Ori     & J05355263$-$0505056 & 5  & 35 & 52.63 & -5  & 5   & 5.63  & G7      & 5530\tablefootmark{a} & \citet{1969ApJ...155..447W} \\
PR Ori       & J05362499$-$0617324 & 5  & 36 & 24.99 & -6  & 17  & 32.55 & K1      & 5170\tablefootmark{a} & \citet{2013ApJ...764..114H} \\
HD 294260    &                   & 5  & 36 & 51.27 & -4  & 25  & 39.97 &         & 6115       & \citet{2004AJ....128.1294C} \\
BE Ori       & J05370010$-$0633273 & 5  & 37 & 0.11  & -6  & 33  & 27.33 & G3      & 5720\tablefootmark{a} & \citet{2012ApJ...752...59H} \\
HBC 502      & J05460788$-$0011568 & 5  & 46 & 7.88  & 0   & -11 & 56.67 & K3      & 4830\tablefootmark{a} & \citet{2008AJ....135..966F} \\
LkH$\alpha$ 310     & J05471098+0019147 & 5  & 47 & 10.98 & 0   & 19  & 14.77 & G6      & 5590\tablefootmark{a} & \citet{2009AA...504..461F}  \\
HD 288313 A  & J05540300+0140217 & 5  & 54 & 3.01  & 1   & 40  & 21.95 & K2 V    & 5040\tablefootmark{a} & \citet{2010AJ....139.1668R} \\
PDS 277      & J08231185$-$3907015 & 8  & 23 & 11.86 & -39 & 7   & 1.62  & F3 Ve   & 6720\tablefootmark{a} & \citet{2003AJ....126.2971V} \\
CR Cha       & J10590699$-$7701404 & 10 & 59 & 6.97  & -77 & 1   & 40.31 &         & 4800       & \citet{2019AA...622A..72V}  \\
Ass ChaT2-21 & J11061540$-$7721567 & 11 & 6  & 15.35 & -77 & 21  & 56.74 & G5 Ve   & 5660\tablefootmark{a} & \citet{2003AJ....126.2971V} \\
DI Cha       &                   & 11 & 7  & 20.72 & -77 & 38  & 7.29  & G2      & 5770\tablefootmark{a} & \citet{2012ApJ...745..119N} \\
CV Cha       & J11122772$-$7644223 & 11 & 12 & 27.71 & -76 & 44  & 22.3  & K0      & 5280\tablefootmark{a} & \citet{2017AA...604A.127M}  \\
Ass ChaT2-54 & J11124268$-$7722230 & 11 & 12 & 42.67 & -77 & 22  & 22.93 &         & 5260       & \citet{2019AA...622A..72V}  \\
HD 135344B   & J15154844$-$3709160 & 15 & 15 & 48.45 & -37 & 9   & 16.03 & F8      & 6640\tablefootmark{a} &   \citet{1995MNRAS.274..977C}                          \\
HT Lup  & J15451286$-$3417305 & 15 & 45 & 12.87 & -34 & 17  & 30.65 & K3 Ve   & 4830\tablefootmark{a} & \citet{2003AJ....126.2971V} \\
HD 142666    & J15564002$-$2201400 & 15 & 56 & 40.02 & -22 & 1   & 40.00 & A8 V    & 7500\tablefootmark{a} & \citet{2003AJ....126.2971V} \\
HD 142527    & J15564188$-$4219232 & 15 & 56 & 41.89 & -42 & 19  & 23.25 &         & 6500       & \citet{2015MNRAS.453..976F} \\
HD 144432    & J16065795$-$2743094 & 16 & 6  & 57.95 & -27 & 43  & 9.76  &         & 7500       & \citet{2015MNRAS.453..976F} \\
Haro 1-6     & J16260302$-$2423360 & 16 & 26 & 3.03  & -24 & 23  & 36.19 & G1      & 5880\tablefootmark{a} & \citet{2014ApJ...786...97H} \\
EM*SR 21  & J16271027$-$2419127 & 16 & 27 & 10.28 & -24 & 19  & 12.62 &         & 5950   & \citet{2003ApJ...584..853P} \\
AK Sco       & J16544485$-$3653185 & 16 & 54 & 44.85 & -36 & 53  & 18.56 &         & 6250       & \citet{2015MNRAS.453..976F} \\
PDS 156      & J18272608$-$0434473 & 18 & 27 & 26.07 & -4  & 34  & 47.46 & G5 III  & 5660\tablefootmark{a} & \citet{2013AA...551A..77T}  \\
DI Cep       & J22561153+5840017 & 22 & 56 & 11.54 & 58  & 40  & 1.77  & G8 IVe  & 5490\tablefootmark{a} & \citet{1997AA...321..497H}  \\
V395 Cep     & J23205208+7414071 & 23 & 20 & 52.12 & 74  & 14  & 7.08  &         & 5470       & \citet{2019AA...622A..72V}  \\
  \hline
\end{tabular}
\tablefoottext{a}{Temperature derived from spectral type using \citet{2013ApJS..208....9P}}
\end{table*}

\begin{table*}
    \caption{Table shows the classification as T-Tau star or HAeBe star to confirm the PMS nature of the star and list weather or not this star has previously been referred to as an intermediate mass PMS star. We find 10 new IMTT stars previously classified as T-Tau stars.  }
    \label{tbl:Star_TTref}
    \small
    \begin{tabular}{lccl}
    \\
    \hline \hline
Name         &  Reference to classification & Known inter-  & Ref. as inter-\\
            & as T-Tau or HAeBe & mediate mass & mediate mass \\
            & & PMS star & PMS star\\
\hline 
BX Ari       & \citet{2014ApJ...786...97H} & Y & \citet{2006RMxAC..26...41P} \\
HBC 338      & \citet{1988cels.book.....H}& Y& \citet{2015AJ....150...95A} \\
LkH$\alpha$ 330 & \citet{2014ApJ...786...97H}& Y& \citet{2016AA...590A..98H} \\
RY Tau       & \citet{2004AJ....128.1294C}& Y& \citet{2004AJ....128.1294C} \\
T Tau        & \citet{2004AJ....128.1294C}& Y& \citet{2004AJ....128.1294C} \\
UX Tau A     & \citet{1999AJ....118.1043H}& Y& \citet{2019ApJ...886..115Y} \\
HQ Tau       & \citet{2014ApJ...786...97H}& Y& \citet{2020AA...642A..99P} \\
HBC 415      & \citet{2014ApJ...786...97H}& Y& \citet{2019ApJ...886..115Y} \\
SU Aur       & \citet{2014ApJ...786...97H}& Y& \citet{2004AJ....128.1294C} \\
HD 34700     & \citet{2005AA...434..671S}& Y& \citet{2020ApJ...888....7L} \\
CO Ori       & \citet{2004AJ....128.1294C}& Y& \citet{2004AJ....128.1294C} \\
HD 35929     & \citet{2018AA...620A.128V}& Y& \citet{2018AA...620A.128V} \\
PDS 115      & \citet{2008MNRAS.387.1335R}& N& \\
GW Ori       & \citet{2004AJ....128.1294C}& Y& \citet{2004AJ....128.1294C} \\
V1650 Ori    & \citet{2019AA...622A..72V}& Y& \citet{2019AA...622A..72V} \\
GX Ori       & \citet{2004AJ....128.1294C}& Y& \citet{2004AJ....128.1294C} \\
RY Ori       & \citet{2019AA...622A..72V}& Y& \citet{2019AA...622A..72V} \\
HBC 442      & \citet{2019AA...622A..72V}& Y& \citet{2019AA...622A..72V} \\
SW Ori       & \citet{1999AJ....118.1043H}& N&\\
V1044 Ori    & \citet{2004AJ....128.1294C}& Y& \citet{2004AJ....128.1294C} \\
EZ Ori       & \citet{2004AJ....128.1294C}& Y& \citet{2004AJ....128.1294C} \\
Brun 252     & \citet{2019AA...622A..72V}& Y& \citet{2019AA...622A..72V} \\
V2149 Ori    & \citet{2019AA...622A..72V}& Y& \citet{2019AA...622A..72V} \\
Brun 555     & \citet{2017ApJ...834..142K}& Y& \citet{1993AJ....105.1087K} \\
Brun 656     & \citet{2016ApJ...818...59D}& N&\\
V815 Ori     & \citet{2016ApJ...818...59D}& Y& \citet{1993AJ....105.1087K} \\
PR Ori       & \citet{2018AJ....156...25R}& N&\\
HD 294260    & \citet{2004AJ....128.1294C}& Y& \citet{2004AJ....128.1294C} \\
BE Ori       & \citet{2016ApJ...818...59D}& N&\\
HBC 502      & \citet{2008AJ....135..966F}& Y&\citet{2002ApJ...569..304M} \\
LkH$\alpha$ 310     & \citet{2008AJ....135..966F}& N&\\
HD 288313 A  & \citet{2010AJ....139.1668R}& N&\\
PDS 277      & \citet{2018AA...620A.128V}& Y& \citet{2018AA...620A.128V} \\
CR Cha       & \citet{2019AA...622A..72V}& Y& \citet{2019AA...622A..72V} \\
Ass ChaT2-21 & \citet{1999AJ....118.1043H}& Y& \citet{2019AA...622A..72V} \\
DI Cha       & \citet{1999AJ....118.1043H}& Y& \citet{2015AA...581A.107M} \\
CV Cha       & \citet{1999AJ....118.1043H}& Y& \citet{2019AA...622A..72V} \\
Ass ChaT2-54 & \citet{2019AA...622A..72V}& Y& \citet{2019AA...622A..72V} \\
HD 135344B   & \citet{2018AA...620A.128V}& Y& \citet{2018AA...620A.128V} \\
HT Lup       & \citet{1999AJ....118.1043H}& N&\\
HD 142666    & \citet{2018AA...620A.128V}& Y& \citet{2019AA...622A..72V} \\
HD 142527    & \citet{2018AA...620A.128V}& Y& \citet{2018AA...620A.128V} \\
HD 144432    & \citet{2018AA...620A.128V}& Y& \citet{2018AA...620A.128V} \\
Haro 1-6     & \citet{2014ApJ...786...97H}& Y& \citet{2015AA...581A.107M} \\
EM*SR 21     & \citet{2014ApJ...786...97H}& Y& \citet{2015AA...581A.107M} \\
AK Sco       & \citet{2018AA...620A.128V}& Y& \citet{2018AA...620A.128V} \\
PDS 156      & \citet{2012AA...548A..79A}& N&\\
DI Cep       & \citet{1999AJ....118.1043H}& N&\\
V395 Cep     & \citet{2019AA...622A..72V}& Y& \citet{2018AA...620A.128V} \\
  \hline
\end{tabular}

\end{table*}

\begin{table*}
    \caption{The stellar parameters and disk types for the sample. The effective temperature is found in literature (see references in table \ref{tbl:Star_Overview}). The distances are \textit{Gaia} DR2 as derived by \citet{2018AJ....156...58B} unless otherwise indicated. The L and E(B-V) are determined by our SED fitting. Mass and age are determined using \citet{2000AA...358..593S} pre-main sequence evolutionary tracks.}
    \label{tbl:Stellarparam}
    \footnotesize

    \begin{tabular}{lllllllll}
    \renewcommand{\arraystretch}{1.2}
    \\
    \hline \hline
Name         & SpT     & $T_{\rm eff}$           & d                & $L_{\star}$                & E(B-V)                   & $M_{\star}$                & Age       & Disk                  \\
             &         & (K)                     & [pc]             & [$L_{\odot}$]               & E(B-V)                   & [$M_{\odot}$]             & [Myr] & Type                    \\
\hline
BX Ari       & K2      & \(5040_{-210}^{+240}\)  & \(238.90_{-3.10}^{+2.97}\)     & \(48.81_{-9.18}^{+3.90}\) & \(4.12_{-0.21}^{+0.01}\) & \(3.82_{-0.79}^{+0.05}\) & \(0.30_{-0.12}^{+0.17}\)  & Ia     \\[2pt]
HBC 338      & G8      & \(5490_{-150}^{+40}\)   & \(290.04_{-6.05}^{+6.32} \)     & \(5.67_{-0.38}^{+0.22} \)  & \(0.79_{-0.03}^{+0.01}\) & \(1.82_{-0.05}^{+0.18}\)   & \(4.19_{-1.08}^{+0.72}\) & Ia   \\[2pt]
LkH$\alpha$ 330     & F7      & \(6240_{-70}^{+100}\)   & \(308.43_{-7.37}^{+7.73} \)      & \(14.39_{-0.92}^{+1.11}\)  & \(0.98_{-0.03}^{+0.03}\) & \(1.93_{-0.07}^{+0.07}\) & \(4.66_{-0.26}^{+0.65}\)  & Ia  \\[2pt]
RY Tau       &         & \(5945_{-143}^{+143}\)  & \(133.51_{-30.10}^{+54.81} \)  \tablefootmark{e} & \(11.97_{-4.20}^{+6.95}\)  & \(0.82_{-0.06}^{+0.01}\) & \(1.95_{-0.17}^{+0.10}\)  & \(4.29_{-0.89}^{+1.43}\) & IIa    \\[2pt]
T Tau        &         & \(5700_{-140}^{+140}\)  & \(143.74_{-1.21}^{+1.22} \)     & \(8.88_{-0.49}^{+1.09} \)  & \(0.70_{-0.03}^{+0.05}\) & \(1.94_{-0.17}^{+0.21}\)   & \(4.01_{-1.09}^{+1.35}\) & I   \\[2pt]
UX Tau A     & G8      & \(5490_{-210}^{+130}\)\tablefootmark{a} & $139.4 \pm1.96$\tablefootmark{b}    & \(8.91_{-2.88}^{+3.11}\)\tablefootmark{b}  &                          & \(2.34_{-0.43}^{+0.29}\)\tablefootmark{b}  & \(1.26_{-0.63}^{+1.03}\)\tablefootmark{c} & Ib  \\[2pt]
HQ Tau       & K0      & \(5280_{-110}^{+60}\)   & \(158.21_{-5.20}^{+5.56} \)      & \(4.63_{-0.32}^{+0.27} \)  & \(1.07_{-0.03}^{+0.02}\) & \(1.85_{-0.06}^{+0.05}\)   & \(3.44_{-0.87}^{+0.62}\) & IIa \tablefootmark{j}   \\[2pt]
HBC 415      & G2      & \(5770_{-50}^{+110}\)   & \(165.18_{-1.30}^{+1.31} \)      & \(7.48_{-0.16}^{+0.40} \)  & \(0.89_{-0.01}^{+0.02}\) & \(1.78_{-0.08}^{+0.07}\)   & \(5.19_{-0.61}^{+0.96}\) & Debris   \\[2pt]
SU Aur       & G4      & \(5680_{-20}^{+40}\)    & \(157.68_{-1.47}^{+1.50} \)      & \(12.75_{-0.34}^{+0.19}\)  & \(0.45_{-0.01}^{+0.00}\) & \(2.22_{-0.05}^{+0.02}\)   & \(2.92_{-0.41}^{+0.19}\) & I    \\[2pt]
HD 34700     & G0 Ive  & \(6060_{-60}^{+80}\)\tablefootmark{i}    & \(353.03_{-4.96}^{+6.18} \)      & \(24.34_{-1.36}^{+1.60}\)  & \(0.07_{-0.02}^{+0.02}\) & \(2.46_{-0.08}^{+0.08}\)   & \(2.41_{-0.35}^{+0.12}\) & I    \\[2pt]
CO Ori       &         & \(6030_{-170}^{+170}\)  & \(399.67_{-6.63}^{+6.85}\)    & \(43.82_{-7.51}^{+2.48}\)  & \(0.67_{-0.07}^{+0.03}\) & \(2.97_{-0.30}^{+0.19}\)   & \(1.48_{-0.49}^{+0.56}\) & IIa    \\[2pt]
HD 35929     &         & \(7000_{-250}^{+250}\)  & \(383.52_{-7.50}^{+7.81} \)      & \(80.78_{-11.91}^{+9.87}\) & \(0.11_{-0.06}^{+0.04}\) & \(3.25_{-0.25}^{+0.22}\)   & \(1.14_{-0.08}^{+0.57}\) & IIa    \\[2pt]
PDS 115      & G2      & \(5770_{-100}^{+100}\)  & \(286.04_{-11.83}^{+12.88} \)     & \(4.27_{-0.37}^{+0.35} \)  & \(0.28_{-0.04}^{+0.02}\) & \(1.48_{-0.08}^{+0.09}\)   & \(8.51_{-1.47}^{+1.25}\) & II   \\[2pt]
GW Ori       &         & \(5700_{-150}^{+150}\)  & \(398.16_{-10.09}^{+10.62} \)     & \(35.47_{-1.53}^{+4.18}\)  & \(0.32_{-0.01}^{+0.05}\) & \(3.02_{-0.15}^{+0.29}\)   & \(1.00_{-0.13}^{+0.45}\) & Ia     \\[2pt]
V1650 Ori    &         & \(6160_{-100}^{+100}\)  & \(342.41_{-5.46}^{+5.63} \)      & \(9.53_{-0.78}^{+0.25} \)  & \(0.18_{-0.03}^{+0.01}\) & \(1.71_{-0.07}^{+0.05}\)   & \(6.20_{-0.41}^{+0.96}\) & IIa    \\[2pt]
GX Ori       &         & \(5410_{-275}^{+275}\)  & \(447.69_{-8.74}^{+9.10} \)     & \(3.10_{-0.51}^{+0.51} \)  & \(0.47_{-0.09}^{+0.08}\) & \(1.53_{-0.24}^{+0.22}\)   & \(6.59_{-3.22}^{+5.51}\) & IIa      \\[2pt]
RY Ori       &         & \(6120_{-110}^{+110}\)  & \(364.79_{-5.11}^{+5.27} \)      & \(9.01_{-0.28}^{+0.96} \)  & \(0.57_{-0.02}^{+0.04}\) & \(1.69_{-0.05}^{+0.11}\)   & \(6.65_{-1.34}^{+0.42}\) & IIa   \\[2pt]
HBC 442      &         & \(6170_{-110}^{+170}\)  & \(381.67_{-5.64}^{+5.81} \)      & \(13.20_{-0.28}^{+1.76}\)  & \(0.13_{-0.00}^{+0.05}\) & \(1.90_{-0.06}^{+0.14}\)   & \(4.75_{-1.14}^{+0.62}\) & IIa    \\[2pt]
SW Ori       & G8      & \(5490_{-210}^{+100}\)  & \(375.68_{-4.35}^{+4.45} \)      & \(3.28_{-0.35}^{+0.26} \)  & \(0.90_{-0.06}^{+0.04}\) & \(1.51_{-0.12}^{+0.17}\)   & \(7.10_{-2.53}^{+2.30}\) & I     \\[2pt]
V1044 Ori    &         & \(5500_{-140}^{+140}\)  & \(387.97_{-5.77}^{+5.94} \)      & \(6.10_{-0.42}^{+0.52} \)  & \(0.22_{-0.04}^{+0.04}\) & \(1.89_{-0.17}^{+0.12}\)   & \(4.00_{-1.12}^{+1.38}\) & Ia      \\[2pt]
EZ Ori       &         & \(5830_{-87}^{+88}\)    & \(399_{-23.4}^{+26.5} \)\tablefootmark{f}     & \(7.54_{-0.85}^{+0.79} \)  & \(0.39_{-0.03}^{+0.03}\) & \(1.75_{-0.09}^{+0.09}\)   & \(5.51_{-0.81}^{+1.06}\) & Ia/IIa   \\[2pt]
Brun 252     &         & \(5890_{-120}^{+120}\)  & \(383.64_{-6.43}^{+6.66} \)      & \(7.16_{-0.38}^{+2.51} \)  & \(0.14_{-0.02}^{+0.13}\) & \(1.68_{-0.09}^{+0.25}\)   & \(6.30_{-2.18}^{+1.24}\) & Debris    \\[2pt]
V2149 Ori    &         & \(6180_{-110}^{+110}\)  & $388\pm5$\tablefootmark{d}     & \(35.80_{-0.91}^{+5.61}\)  & \(0.81_{-0.01}^{+0.06}\) & \(2.68_{-0.06}^{+0.22}\)   & \(2.04_{-0.52}^{+0.07}\) & Ib/IIb     \\[2pt]
Brun 555     & K2      & \(5040_{-50}^{+240}\)   & \(453.24_{-8.84}^{+9.19} \)      & \(11.37_{-0.40}^{+1.03}\)  & \(0.21_{-0.01}^{+0.05}\) & \(2.49_{-0.11}^{+0.11}\)   & \(0.87_{-0.16}^{+0.65}\) & Debris \\[2pt]
Brun 656     & G2 III  & \(5770_{-90}^{+150}\)   & \(466.80_{-9.78}^{+10.19} \)     & \(27.40_{-1.32}^{+3.62}\)  & \(0.50_{-0.02}^{+0.05}\) & \(2.72_{-0.15}^{+0.21}\)   & \(1.53_{-0.18}^{+0.48}\)  & Debris  \\[2pt]
V815 Ori     & G7      & \(5530_{-40}^{+60}\)    & \(397.45_{-4.78}^{+4.86} \)      & \(5.46_{-0.14}^{+0.09} \)  & \(0.25_{-0.01}^{+0.00}\) & \(1.76_{-0.05}^{+0.06}\)   & \(4.82_{-0.50}^{+0.45}\) & Ib   \\[2pt]
PR Ori       & K1      & \(5170_{-130}^{+110}\)  & \(408_{-4}^{+4} \) \tablefootmark{g}     & \(10.83_{-0.69}^{+1.11} \)  & \(0.45_{-0.04}^{+0.05}\) & \(2.50_{-0.31}^{+0.03}\)   & \(1.05_{-0.22}^{+0.75}\) & Ia/IIa    \\[2pt]
HD 294260    &         & \(6115_{-168}^{+168}\)  & \(406.70_{-6.74}^{+6.96}\)    & \(8.74_{-0.53}^{+0.78} \)  & \(0.16_{-0.02}^{+0.04}\) & \(1.68_{-0.07}^{+0.13}\)   & \(6.72_{-1.54}^{+0.84}\) & IIa    \\[2pt]
BE Ori       & G3      & \(5720_{-60}^{+160}\)   & \(392.27_{-5.29}^{+5.43} \)      & \(7.95_{-0.21}^{+0.80} \)  & \(1.41_{-0.01}^{+0.05}\) & \(1.86_{-0.09}^{+0.10}\)   & \(4.55_{-0.70}^{+1.13}\) & IIa    \\[2pt]
HBC 502      & K3      & \(4830_{-230}^{+210}\)  & \(410.94_{-8.32}^{+8.66} \)      & \(10.71_{-0.98}^{+1.26}\)  & \(1.53_{-0.06}^{+0.07}\) & \(2.01_{-0.56}^{+0.52}\)   & \(0.53_{-0.24}^{+0.46}\) & Ib \\[2pt]
LkH$\alpha$ 310     & G6      & \(5590_{-100}^{+90}\)   & \(422.46_{-11.25}^{+11.87} \)     & \(6.80_{-0.52}^{+0.48} \)  & \(1.16_{-0.02}^{+0.01}\) & \(1.87_{-0.12}^{+0.13}\)   & \(4.25_{-0.87}^{+0.97}\) & Ia   \\[2pt]
HD 288313 A  & K2 V    & \(5040_{-440}^{+300}\)  & \(418_{-17}^{+17}\)\tablefootmark{h}   & \(38.47_{-13.20}^{+10.00}\) & \(0.34_{-0.21}^{+0.11}\) & \(3.47_{-1.59}^{+0.23}\)   & \(0.52_{-0.39}^{+1.32}\) & - \\[2pt]
PDS 277      & F3 Ve   & \(6720_{-80}^{+90}\)    & \(342.57_{-4.20}^{+4.31} \)      & \(9.68_{-0.67}^{+0.61} \)  & \(0.03_{-0.02}^{+0.02}\) & \(1.59_{-0.09}^{+0.05}\)   & \(8.66_{-1.12}^{+0.33}\) & I    \\[2pt]
CR Cha       &         & \(4800_{-230}^{+230}\)  & \(186.47_{-0.79}^{+0.81} \)      & \(3.72_{-0.49}^{+0.57} \)  & \(0.40_{-0.09}^{+0.10}\) & \(1.62_{-0.39}^{+0.28}\)   & \(1.51_{-0.80}^{+1.59}\) & IIa \\[2pt]
Ass ChaT2-21 & G5 Ve   & \(5660_{-70}^{+20}\)    & \(164.77_{-3.77}^{+3.95} \)     & \(10.37_{-0.63}^{+0.59}\)  & \(1.03_{-0.03}^{+0.00}\) & \(2.09_{-0.06}^{+0.09}\)   & \(3.23_{-0.36}^{+0.17}\) & Debris   \\[2pt]
DI Cha       & G2      & \(5770_{-50}^{+110}\)   & \(189.58_{-1.06}^{+1.08} \)      & \(9.82_{-0.33}^{+0.61} \)  & \(0.73_{-0.01}^{+0.03}\) & \(1.95_{-0.10}^{+0.08}\)   & \(4.08_{-0.66}^{+0.75}\) & IIa   \\[2pt]
CV Cha       & K0      & \(5280_{-110}^{+60}\)   & \(192.17_{-0.97}^{+0.98} \)     & \(4.63_{-0.15}^{+0.19} \)  & \(0.46_{-0.02}^{+0.02}\) & \(1.85_{-0.05}^{+0.05}\)   & \(3.44_{-0.87}^{+0.51}\) & IIa    \\[2pt]
Ass ChaT2-54 &         & \(5260_{-200}^{+200}\)  & \(202.92_{-16.18}^{+19.19} \)     & \(5.03_{-0.99}^{+0.71} \)  & \(0.55_{-0.08}^{+0.03}\) & \(1.92_{-0.24}^{+0.08}\)   & \(3.06_{-1.16}^{+2.24}\)  & Debris   \\[2pt]
HD 135344B   & F4      & \(6640_{-80}^{+130}\)   & \(135.26_{-1.40}^{+1.44} \)      & \(8.09_{-0.24}^{+0.16} \)  & \(0.21_{-0.01}^{+0.01}\) & \(1.50_{-0.01}^{+0.02}\)   & \(9.07_{-0.25}^{+0.80}\) & Ib \tablefootmark{k}    \\[2pt]
HT Lup  & K3 Ve   & \(4830_{-230}^{+210}\)  & \(153.53_{-1.42}^{+1.45} \)      & \(5.95_{-0.73}^{+0.69} \)  & \(0.48_{-0.09}^{+0.08}\) & \(1.80_{-0.50}^{+0.33}\)   & \(0.95_{-0.47}^{+0.89}\) & IIa \\[2pt]
HD 142666    & A8 V    & \(7500_{-280}^{+500}\)  & \(147.66_{-1.15}^{+1.17} \)      & \(11.81_{-1.63}^{+1.93}\)  & \(0.32_{-0.06}^{+0.05}\) & \(1.62_{-0.03}^{+0.13}\)   & \(8.37_{-1.22}^{+0.83}\) & IIa    \\[2pt]
HD 142527    &         & \(6500_{-250}^{+250}\)  & \(156.63_{-1.16}^{+1.18} \)      & \(19.31_{-3.68}^{+1.29}\)  & \(0.30_{-0.08}^{+0.03}\) & \(2.06_{-0.20}^{+0.18}\)   & \(3.80_{-1.07}^{+1.80}\) & Ia    \\[2pt]
HD 144432    &         & \(7500_{-280}^{+500}\)  & \(154.68_{-1.39}^{+1.40} \)      & \(12.50_{-1.26}^{+4.06}\)  & \(0.10_{-0.04}^{+0.09}\) & \(1.64_{-0.01}^{+0.19}\)   & \(8.14_{-1.95}^{+0.63}\) & IIa   \\[2pt]
Haro 1-6     & G1      & \(5880_{-110}^{+40}\)   & \(133.81_{-1.19}^{+1.21} \)      & \(12.78_{-0.19}^{+0.16}\)  & \(1.39_{-0.00}^{+0.00}\) & \(2.04_{-0.04}^{+0.09}\)   & \(3.50_{-0.28}^{+0.11}\) & Ib    \\[2pt]
EM*SR 21     &         & \(5950_{-300}^{+300}\)  & \(137.86_{-1.07}^{+1.1} \)      & \(7.03_{-1.54}^{+2.01} \)  & \(1.25_{-0.07}^{+0.07}\) & \(1.64_{-0.22}^{+0.36}\)   & \(6.71_{-3.01}^{+3.59}\) & Ib   \\[2pt]
AK Sco       &         & \(6250_{-250}^{+250}\)  & \(140.04_{-1.21}^{+1.24} \)      & \(6.94_{-1.51}^{+0.68} \)  & \(0.28_{-0.09}^{+0.04}\) & \(1.52_{-0.13}^{+0.45}\)   & \(8.57_{-1.91}^{+3.73}\) & Ia       \\[2pt]
PDS 156      & G5 III  & \(5660_{-70}^{+20}\)    & \(397.59_{-6.44}^{+6.64} \)      & \(21.12_{-0.72}^{+0.50}\)  & \(1.03_{-0.03}^{+0.00}\) & \(2.61_{-0.04}^{+0.08}\)   & \(1.80_{-0.15}^{+0.07}\) & II   \\[2pt]
DI Cep       & G8 IVe  & \(5490_{-50}^{+110}\)   & \(430.02_{-5.78}^{+5.94} \)      & \(9.82_{-0.37}^{+0.63} \)  & \(0.73_{-0.01}^{+0.03}\) & \(2.20_{-0.12}^{+0.09}\)   & \(2.55_{-0.34}^{+0.63}\) & IIa    \\[2pt]
V395 Cep     &         & \(5470_{-70}^{+70}\)    & \(188.41_{-1.09}^{+1.11} \)      & \(4.71_{-0.24}^{+0.20} \)  & \(0.27_{-0.02}^{+0.02}\) & \(1.73_{-0.08}^{+0.09}\)   & \(4.86_{-0.79}^{+0.9}\) & Ia     \\[2pt]

  \hline
\end{tabular}\\
\tablefoottext{a}{Value from Espaillat et al. 2010}.
\tablefoottext{b}{Value from Akeson et al. 2019}.
\tablefoottext{c}{Value from Kraus \& Hilldebrand 2009}.
\tablefoottext{d}{Value from Kounkel et al. 2016}.
\tablefoottext{e}{Value from \citet{1997A&A...323L..49P}}.
\tablefoottext{f}{\citet{2018AJ....156...84K}}.
\tablefoottext{g}{\citet{2019AA...624A...6Y}}
\tablefoottext{h}{\citet{2020AA...633A..51Z}}.
\tablefoottext{i} $T_{eff}$ increased see footnote 1 in section 2.5 for explanation.
\tablefoottext{j}{10$\mu$m feature shown in \citet{2009ApJS..180...84W}}
\tablefoottext{K}{Absence of 10$\mu$m feature shown in \citet{2008AA...491..809F}}
\end{table*}
    
\begin{figure*}[ht]
        \includegraphics[width=\textwidth]{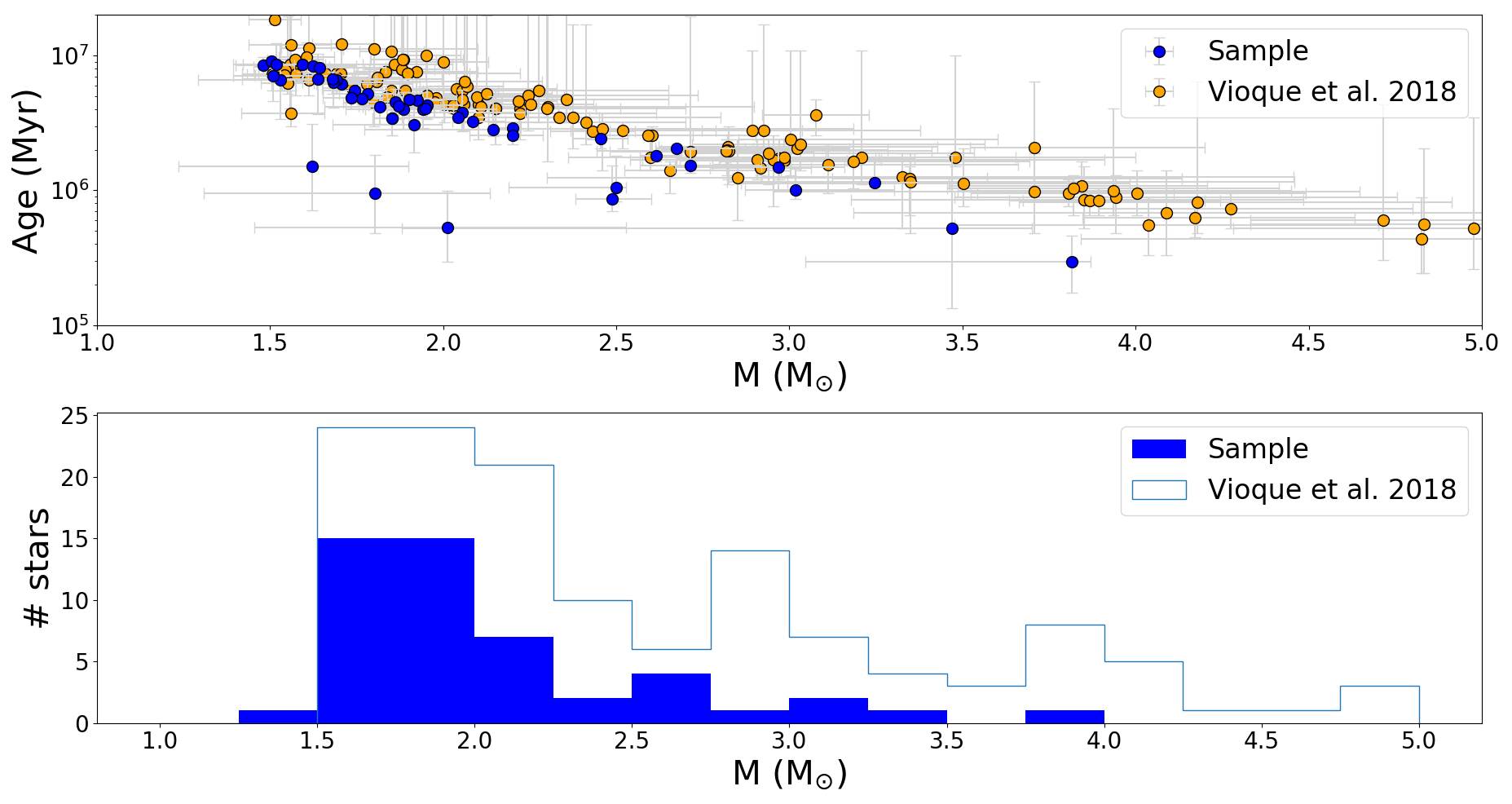}
        \caption{Comparison between the sample of IMTT stars in this paper with the HAeBe stars sample  from \citet{2018AA...620A.128V}. Only stars up to 5$M_{\odot}$ are shown for the HAeBe stars in these panels. \textbf{Upper panel} shows the age-mass relation among the IMTT star sample in comparison with the Herbig star sample, the IMTT star sample is with each mass bin on average younger than the HAeBe stars. \textbf{Lower panel} shows the mass distribution in the IMTT star sample compared to the mass distribution among the HAeBe stars.}
        \label{fig:C-overview}
    \end{figure*}

    The majority of the stars have masses $\leq 2M_{\odot}$, only 4 stars have masses $\geq 3M_{\odot}$ (figure \ref{fig:C-overview} lower panel) in contrast with the sample of \citet{2018AA...620A.128V}. This is consistent with the fact that stars of higher mass are rarer, and also that higher mass stars evolve faster into A and B stars and leave our predefined temperature range in the HR-diagram in a shorter time. In addition the higher mass stars in the F-K temperature region may not yet have cleared their envelope and may still be embedded. They therefore fall outside our optical selection criteria and do not show in our sample.

    Compared with the \citet{2018AA...620A.128V} sample the stellar age in each mass bin is on average younger (figure \ref{fig:C-overview} upper panel). This means that for example if we look at stars with mass between 1.5-2.0 $M_{\odot}$ the average age is lower than the typical age of the Herbig stars. This shows that we indeed have possible candidate progenitors of the Herbig stars in the sample. As expected, the stellar age increases with lower mass, because PMS evolutionary timescales decrease with increasing stellar mass. The median age for the sample is ~4 Myr.
    
    With the help of the fitted photospheric spectrum the infrared excess can be determined. Infrared, millimeter and sub-millimeter photometry were collected for this sample from the \textit{AllWISE}, \textit{AKARI}, \textit{Spitzer} and \textit{Herschel} point source catalogs but also from other publications (see table \ref{tbl:phot_ref} on details for each source star) to build up the SEDs so that the disk can be analysed. In gas rich disks we expect to see an near infrared excess in the SED in contrast to debris disks where the infrared excess first becomes visible beyond 10 $\mu$m. In the sample we can identify 6 disks where the SED resembles that of a debris disk (HBC 415, Brun 252, Brun 555, Brun 656, Ass Cha T2-21, Ass Cha T2-54). Since we are interested in gas rich disks these stars remain in our presented sample of sources but are removed from the disk analysis.
\begin{table*}
  \caption{References to photometry used for the SED. The last column contains the \textit{AOR}-key from the CASSIS database to the \textit{Spitzer} spectra used for the source.
         \textbf{Optical [0.38-1.24 $\mu$m]:} 1) \citet{2016MNRAS.463.4210N}, 2) \citet{2015AAS...22533616H}, 3) \citet{2018AA...616A...1G}, 4) \citet{2011MNRAS.411..435B}, 5) \citet{1978AAS...34..477M}, 6) \citet{2017yCat.4034....0H}, 7) \citet{2016KPCB...32..260A}, 8) \citet{2015yCat.5145....0M}, 9) \citet{2018AA...620A.128V}, 10) \citet{2006ApJ...653..657M}, 11) \citet{2017KPCB...33..250Y}, 12) \citet{2008AJ....136..735L}, 13) \citet{2012ApJ...752...59H}, 14) \citet{2006ApJ...638.1004A}, 15) \citet{2019AA...623A..72K}, 16) \citet{2013ApJ...764..114H}, 17) \citet{2018AJ....155...39O}, 18) \citet{2017MNRAS.471..770M}, 19) \citet{2011AJ....142...15G}, 20) \citet{2003AA...404..913S}, 21) \citet{2018AA...612A..96F}, 22) \citet{2017AJ....153...75K}. \textbf{[1.25-2.23 $\mu$m]:} 1) \citet{2014yCat.2328....0C}, 2) \citet{2003yCat.2246....0C}. \textbf{[2.24-60 $\mu$m]:} 1) \citet{2014yCat.2328....0C}, 2) \citet{2015AC....10...99A}, 3) \citet{2010AA...514A...1I}, 4) \citet{2003PASP..115..965E}, 5) \citet{2012yCat.2311....0C} 6) \citet{2014ApJ...784..126E}, 7) \citet{2010ApJS..186..259R}, 8) \citet{2019AJ....158...54E}, 9) \citet{2012AJ....144..192M}, 10) \citet{2013ApJ...768...99P}, 11) \citet{2017ApJS..229...28G}, 12) \citet{2013ApJS..207....5F}, 13) \citet{2008AJ....135..966F}, 14) \citet{2008ApJ...684..654L}, 15) \citet{2008ApJ...675.1375L}, 16) \citet{2015ApJS..220...11D}, 17) \citet{2008ApJS..177..551M}, 18) \citet{1988iras....7.....H}, 19) \citet{2017ApJ...836...34M}. \textbf{[61-500 $\mu$m]:} 1) \citet{2003PASP..115..965E}, 2) \citet{2020yCat.8106....0H}, 3) \citet{2017ApJ...849...63R}, 4) \citet{2015AC....10...99A}, 5) \citet{2010ApJ...724..835W}, 6) \citet{2010ApJS..186..259R}, 7) \citet{2011ApJ...727...26S}, 8) \citet{2008ApJS..177..551M}, 9)\citet{2018AA...619A..52B}, 10) \citet{2013ApJ...775...63D}, 11) \citet{2015ApJS..220...11D}, 12) \citet{2015AA...581A..30R}. \textbf{[$>$500 $\mu$m]}: 1) \citet{2013ApJ...771..129A}, 2) \citet{2013ApJ...773..168M}, 3) \citet{2015ApJ...801...91D}, 4) \citet{2010ApJ...709L.114D}, 5) \citet{2012ApJ...751..115H}, 6) \citet{2018ApJ...859...33G}, 7) \citet{2014ApJ...790...49K}, 8) \citet{2017ApJ...834..142K}, 9) \citet{2016ApJ...831..125P}, 10) \citet{2011ApJ...727...26S}, 11) \citet{2018ApJ...859...21A}, 12) \citet{2008ApJ...686L.115C}, 13) \citet{2013ApJ...775...63D}.}
\label{tbl:phot_ref}
\begin{tabular}{lllllll}
\hline \hline
 & 0.38-1.24 & 1.25-2.23 & 2.23-60 & 61-500 & >500 & CASSIS  \\
 Name & $\mu$m & $\mu$m & $\mu$m & $\mu$m & $\mu$m & \textit{AOR}-key \\

\hline
BX Ari       & 1,2            & 1 & 1,2,3    &          &             & 14975232\\
HBC 338      & 1,3            & 1 & 1,2,4    & 1,2      &             & 21868544\\
LkH$\alpha$ 330     & 1,2            & 1 & 1,2,3,4  & 1,2      &      & 56344816\\
RY Tau       & 1,2,4          & 2 & 2,3,4,5  & 3,4      & 1,2,3       & 26141184\\
T Tau        & 1,2,3,5,6      & 2 & 1,2,3    & 2,4      & 1,2,3,4,5,6 &\\
UX Tau A     & 1,2,6,7        & 2 & 1,2,4    & 3,4,5    & 1,2,5       & 26140928\\
HQ Tau       & 1,6            & 1 & 1,2,3,6  & 3,6      & 1           & 27057664\\
HBC 415      & 1,3            & 2 & 1,7      &          & 3,6         & 3543040\\
SU Aur       & 1,5            & 2 & 1,2,3,8  & 2        & 1,2         & 27066880\\
HD 34700     & 1,2,3,8,9      & 2 & 2,5      & 4        &             &\\
CO Ori       & 1,10           & 2 & 1,2,3    &          &             & 21870336\\
HD 35929     & 1,2,3,10,14    & 2 & 1,2,3    &          &             & 10998528\\
PDS 115      & 1,3            & 2 & 1,2,3    &          &             &\\
GW Ori       & 1,2,3,5        & 2 & 1,2,3    &          &             & 21870592\\
V1650 Ori    & 1,2,3          & 2 & 1,2,3    &          &             & 21870848\\
GX Ori       & 1,2,3          & 2 & 1,2,3    &          &             & 21871360\\
RY Ori       & 1,2,3          & 2 & 1,2,3    &          &             & 21871616\\
HBC 442      & 1,2,5,10       & 2 & 1,2,3,9  &          &             & 18832640\\
SW Ori       & 3,11           & 2 & 1,2,3,10 & 2        &             &\\
V1044 Ori    & 1,2,3,5        & 2 & 2,3,9    & 2        &             & 21872640\\
EZ Ori       & 1,2,3,5        & 2 & 1,2,3,9  & 2        &             & 21872896\\
Brun 252     & 1,2,3,14       & 2 & 1,       &          &             &\\
V2149 Ori    & 1,2,3          & 1 & 1,9      &          & 7           & 18802176\\
Brun 555     & 1,2,3,5        & 2 & 1,11     &          & 7,8         & 21874176\\
Brun 656     & 1,2,3,5        & 2 & 1,11     &          & 7,8         & 21875456\\
V815 Ori     & 1,2,3,5,11     & 2 & 1,11     &          &             & 21877504\\
PR Ori       & 1,2,3,16       & 2 & 1,12     & 2        & 7,8         & 18806784\\
HD 294260    & 1,2,3          & 2 & 1,9      & 2        &             & 21878528\\
BE Ori       & 2,3            & 2 & 1,2,3,12 & 2        &             & 18815232\\
HBC 502      & 3              & 2 & 1,11,13  & 2        &             & 12642048\\
LkH$\alpha$ 310     & 3              & 2 & 1,9      &          &             & 18756608\\
HD 288313 A  & 1,2,3,17       & 1 & 1,       & 2        & 7           &\\
PDS 277      & 1,2,3,12,18    & 1 & 1,2      & 2        &             &\\
CR Cha       & 1,2,3,19       & 2 & 1,2,3,14 & 2        & 9           & 26143744\\
Ass ChaT2-21 & 1,2,3,19       & 1 & 1,2,3,15 &          &             & 12696320\\
DI Cha       & 1,2,3,14,19    & 2 & 1,2,3,16 & 2,11     & 9           & 12697345\\
CV Cha       & 1,2,3,4        & 2 & 1,2,16   & 3,11     &             & 12697088\\
Ass ChaT2-54 & 1,2,3,19       & 2 & 1,16     & 2,4      &             & 12695552\\
HD 135344B   & 1,3,19         & 2 & 1,2,3    & 1,2,7    & 10          & 56557088\\
HT Lup       & 1,2,3,4,19     & 2 & 1,3,17   & 8,9      & 11          & 22806016\\
HD 142666    & 1,3,6,10,12,19 & 2 & 1,2,3    & 2,4      &             &\citet{2010ApJ...721..431J}\\
HD 142527    & 3,4,10,19,20   & 2 & 1,18     & 2        &             & \citet{2010ApJ...721..431J}\\
HD 144432    & 1,3,10,20      & 2 & 1,2,3    & 4        &             &\citet{2010ApJ...721..431J}\\
Haro 1-6     & 2,3,6          & 2 & 1,3,19   & 10,11,12 & 12,13       & 12698368\\
EM*SR 21     & 2,3,6          & 2 & 1,2,3    & 1,3      & 2           & 12698880\\
AK Sco       & 1,4,9,10,19    & 2 & 1,2,3,20 &          &             & 12700160\\
PDS 156      & 1,2,3          & 2 & 1,2      &          &             &\\
DI Cep       & 1,2,3          & 2 & 1,2,3    &          &             & 21887232\\
V395 Cep     & 2,3,21         & 2 & 1,2,3    & 4        &             & 21887744\\
\hline

\end{tabular}
\end{table*}    

\section{Qualitative disk analysis}

    In this section, we examine the SED and use archived low resolution mid-infrared spectra from the Combined Atlas of Sources with Spitzer Infrared Spectra (CASSIS) \citep{2011ApJS..196....8L} to perform a comparison with the HAeBe and T-Tauri population in a qualitative analysis of the disks (see table \ref{tbl:phot_ref} for CASSIS \textit{AOR}-key). The spectra of three stars, HD 142666, HD 142527 and HD 144432, have been taken from \citet{2010ApJ...721..431J}. By using the spectral slope between 13 and 30 $\mu$m we classify the disks into disk geometries and compare the distribution with samples of HAeBe stars. We obtain the strength of the 10 $\mu$m silicate feature from archived low-resolution Spitzer spectra to examine the state of the silicate grain evolution in the disks with those around HAeBe stars. We inspect the mid-infrared spectra for evidence of PAH emission and compare the detection frequency with the HAeBe and lower mass T-Tauri stars. We perform a comparison between the spectral slope and the near infrared excess in order to examine the inner disk opacity. All values measured in the Spitzer spectra for the sources are available in table \ref{tbl:Flux_mesurements}.

\subsection{Group I vs. Group II}
    Spatially resolved observations in scattered light, infrared and submm wavelengths have given rise to the Meeus Group I and Group II classification \citep{2001AA...365..476M} to represent two different disk geometries \citep[e.g. see][]{2018AA...620A..94G}. A similar approach of classifying disks in terms of disk-geometry (full-, pre-tansitional and transitional disk) based on their SED, has also been used in T Tauri literature  \citep{2012ApJ...758...31L,2014ApJ...784..126E,2014prpl.conf..497E}. Ideally, imaging data is used to establish the geometry of the disks in our sample. Unfortunately such data are not available for all stars. We can nevertheless use the Group I/Group II classification to infer the disk geometry for our sample. In Appendix C we verify if our classification in transitional disks (Group I) and self-shadowed/compact (Group II) is supported by published imaging data. We find that for 16 of 18 sources, where spatially resolved image data is available, our classification is supported.  

    In Group I disks the upper layers of the (flared) disk at spatial scales of tens of AU are directly irradiated by the central star. Almost all Group I disks have a large gap where the dust has been cleared out. That allows the central star to irradiate the outer disk \citep{2016AA...587A..62K}. The inner rim of the outer disk scatters stellar radiation and significantly contributes to the SED in the mid- to far-infrared \citep{2012ApJ...752..143H}. In Group II disks the inner rim of the disk cast a shadow on the outer disk \citep{2004AA...417..159D}. The outer disk, if present, does not receive direct stellar photons and therefore usually shows no or weak emission in scattered light \citep{2017AA...603A..21G}, and a 'blue' mid- to far-IR SED. Other effects that can strongly influence the SED are the disk outer radius \citep[a small outer radius results in a Group II SED, see][]{2003AA...398..607D}, or a misalignment between the inner and the outer disk \citep{2015ApJ...798L..44M}.

    We use the method of \citet{2009AA...502L..17A} to classify the disks into Group I and Group II objects by based on the infrared slope of the continuum between 13\,${\mu}m$ and 30\,${\mu}$m by using the Spitzer spectra obtained from CASSIS, averaging the flux values in a 0.2 ${\mu}m$ wide window around 13 ${\mu}m$ and 30 ${\mu}m$ separately. The slope of the spectrum is defined as the ratio between the fluxes, $[F_{30}/F_{13}]$. We define the dividing line between the two groups to be at 2.1 suggested by \citet{2016AA...587A..62K} were Group I disks are identified with a flux-ratio $[F_{30}/F_{13}] \geq 2.1$.

    For 7 sources without Spitzer-spectra at 13 and 30 ${\mu}m$ the classification has instead been based on the photometric infrared excess at 60 ${\mu}m$, $E_{60}$, \citet{2009AA...502L..17A} suggested that for HAeBe stars there is a strong correlation between the disk geometries as determined by the 60 $\mu$m excess and the 13 to 30 $\mu$m spectral slope. Therefore we make use of a method used by \citet{2006AA...457..171A} where they for HAeBe stars defined a Group I source having an excess of $E_{60}$ $\geq 10$ mag. However this limit was set for stars of typical $T_{\rm eff}$ of $\sim8000-10000$K. The stars in our sample are cooler (and therefore redder). Since this leads to less excess (see also appendix A) the 10 magnitude excess limit set for the Herbig stars needs to be adjusted. Using BC for PMS stars produced by \citet{2013ApJS..208....9P} we see that there is a difference in bolometric correction of $BC_J \sim 0.75$ between HAeBE and IMTTS stars. We therefore use the excess limit $E_{60} \geq 9.25$ mag for Group I disks around the cooler IMTT stars. We measured the flux above the photosphere at 60 $\mu$m and used the fitted model spectrum to calculate the excess expressed in magnitudes.

    We find that 20 sources fall into the Group I category, 19 sources in the Group II category. In addition we have 3 stars that fall in between Group I and Group II due to the uncertainty in their spectral index and 1 source, HD 288313A, where the lack of data makes it impossible using any of these direct methods to determine the Group membership. This result can be compared to \citep{2010ApJ...721..431J} where 53 HAeBe stars were categorized into the two Meeus groups using the classification of \citep{2005AA...437..189V}. In comparison, the HAeBe disk distribution between Group I and Group II are 20 and 33, respectively. 

\subsection{Inner disk and disk flaring}

    \citet{2009AA...502L..17A} discovered a correlation between the excess at 7 $\mu$m and the spectral slope between 13 and 30 $\mu$m. This relationship was found for both Group I and Group II disks. The correlation was interpreted as a result of an inner disk casting a shadow on the outer disk. A larger near infrared excess implies a higher inner scale height, resulting in a more shadowed disk with a 'bluer' spectral slope. We wish to investigate if such a correlation is also present in the IMTT star sample.

\begin{figure*}[ht]
        \includegraphics[width=\textwidth]{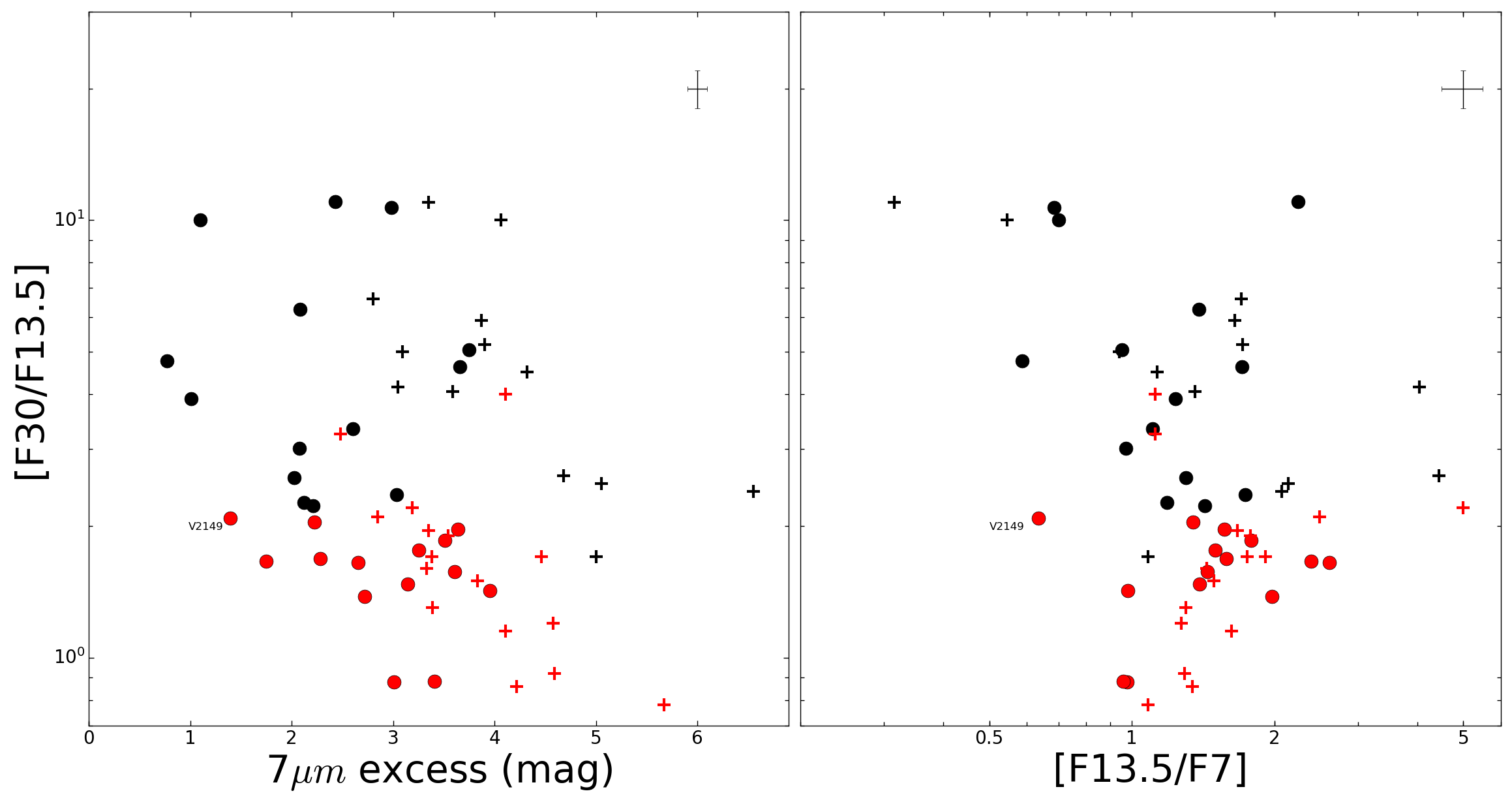}
        \caption{The crosses are the HAeBe sample from \citet{2009AA...502L..17A} and the solid circles are the IMTT stars in our sample. Black indicates a Group I source and red a Group II source. The average uncertainty in the measurement is displayed in the top right corner of the panel. \textbf{Left panel:} The relationship between the spectral slope between 13 and 30 $\mu$m and the 7 $\mu$m excess. Compared to the HAeBe stars there is a lack of IMTT stars with a 7 $\mu$m excess above 4 magnitudes. \textbf{Right panel:} The color comparison with the Herbig stars reveal that the distribution is similar to the HAeBe population. The location of V2149 Ori is marked.}
        \label{fig:C-EXC}
    \end{figure*}

    We therefore measure the excess at 7 $\mu$m by subtracting the  fitted Kurucz model flux from the observed SED at 7 $\mu$m and then calculate the difference expressed in magnitudes. The results are compared with the HAeBe stars from \citet{2009AA...502L..17A} and can be seen in Fig~\ref{fig:C-EXC} (left panel).
    
    The IMTT stars occupy a similar region in figure \ref{fig:C-EXC} (left panel). The trend that stars with a larger near infrared excess have a lower [$F_{30}/F_{13.5}$] index also holds for the IMTT stars. There is one difference: the maximum 7 $\mu$m excess is 4.5 mag for IMTT stars and 6 mag for HAeBe stars. In appendix A we show that this difference can simply be explained by the changes in luminosity and temperature of an evolving intermediate mass PMS star, without the need for differences in inner disk structure.

    The correlation between the [$F_{30}/F_{13.5}$] and [$F_{13.5}/F_{7}$] flux ratios described by \citet{2009AA...502L..17A} for the Group II disks in HAeBe stars is also present among the IMTT stars (figure~\ref{fig:C-EXC}, right panel). The Group II disks are concentrated in a similar fashion in color and there is less of a spread among the Group I disks in the [$F_{13.5}/F_{7}$] ratio then for the HAeBe stars. There is one exception which is the Group II disk of V2149 Ori which is redder, with a lower [$F_{13.5}/F_{7}$] flux ratio and lies somewhat outside towards lower 7 $\mu$m excess the rest of the Group II concentration and has been marked in both panels in figure \ref{fig:C-EXC}.

\subsection{10$\mu$m silicate feature}
    The bulk of dust in protoplanetary disks is made up of silicate grains. The 10 $\mu$m silicate feature is sensitive to the size of the dust grains, the chemical composition, and the lattice structure in the grains. The shape and strength of the feature can be used as a signpost of dust processing in the disk \citep{2003AA...400L..21V}. 

    The shape of the silicate emission feature can be measured by the ratio of the flux at 11.3 $\mu$m and 9.8 $\mu$m, [$F_{11.3}/F_{9.8}$] and the peak strength of the feature by the continuum divided flux at 9.8 $\mu$m see \citep[see][]{2003AA...400L..21V}. As small silicate grains (0.1 ${\mu}m$) are removed from the inner disk, primarily by grain growth, the strength of the 10 ${\mu}m$ band weakens. The evolution also leads to a flatter band shape leading to a higher [$F_{11.3}/F_{9.8}$] \citep{2003AA...400L..21V, 2003AA...412L..43P}. The ratio is also sensitive to the presence of a strong crystalline olivine emission band at 11.3 $\mu$m.

\begin{figure}[ht]
        \includegraphics[width=8.8cm]{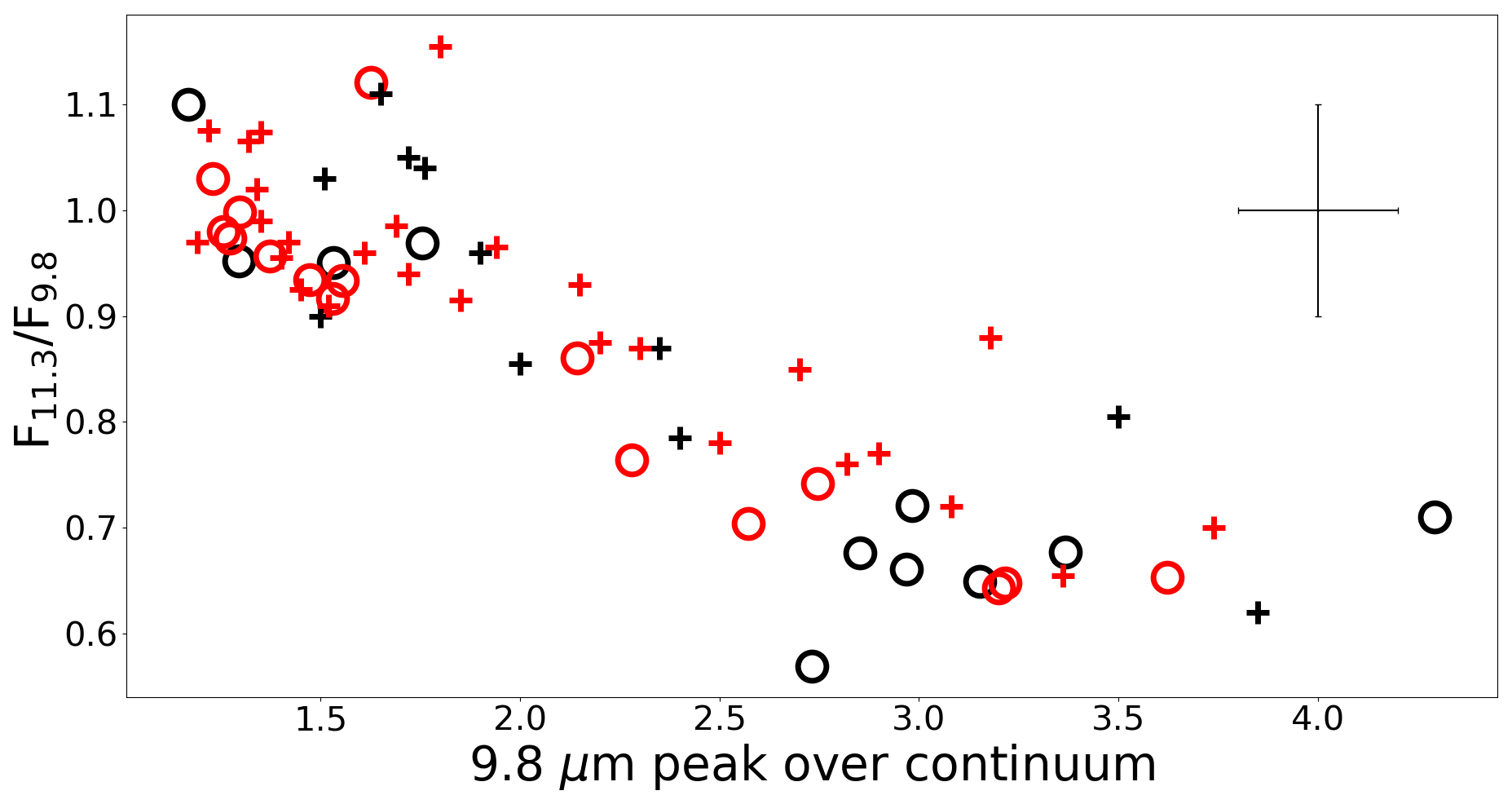}
        \caption{The degree of processing for the silicate grains in the disks around the IMTT star sample (open circles) and for the HAeBe sample from \citet{2010ApJ...721..431J} (crosses). Black are Group I disks and red are Group II disks. Smaller grains leads to a high peak over continuum intensity and a low [$F_{11.3}/F_{9.8}$] ratio. As grains grow the [$F_{11.3}/F_{9.8}$] ratio increases and the peak over continuum intensity weakens.}
        \label{fig:C-grainevol}
    \end{figure}

\begin{figure*}[ht]
        \includegraphics[width=\textwidth]{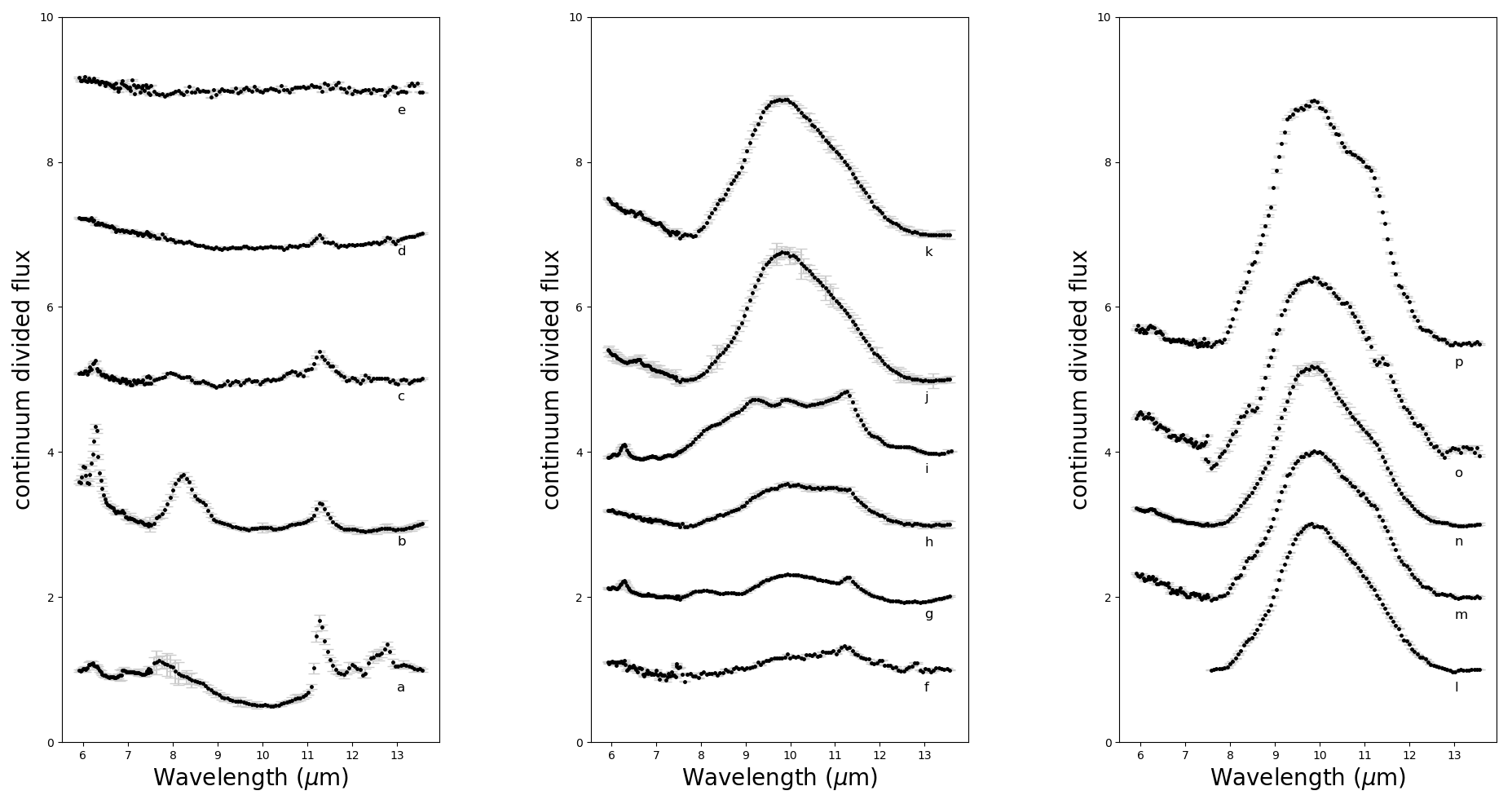}
        \caption{This shows the continuum divided normalized peak strength over the 10 $\mu$m feature for those sources classified as Group I. The continuum between 7.5 and 13.5 $\mu$m has been normalized to 1. The leftmost panel shows the spectra of sources were no detection of a silicate feature is found: a) Haro 1-6, b) EM*SR 21, c) V815 Ori, d) UX Tau e) HBC 502. From the second left panel the disks with detected silicate feature is shown in ascending order of peak strength: f) HBC 338, g) LkH$\alpha$ 330, h) V395 Cep, i) HD 142527, j) GW Ori, k) DI Cep, l) BX Ari, m) EZ Ori, n) AK Sco, o) LkH$\alpha$ 310, p) V1044 Ori. The scale is indicated in the leftmost panel and is the same for all sources, spectra are shifted to allow comparison.}
        \label{fig:C-GIsil}
    \end{figure*}

\begin{figure*}[ht]    
            \includegraphics[width=\textwidth]{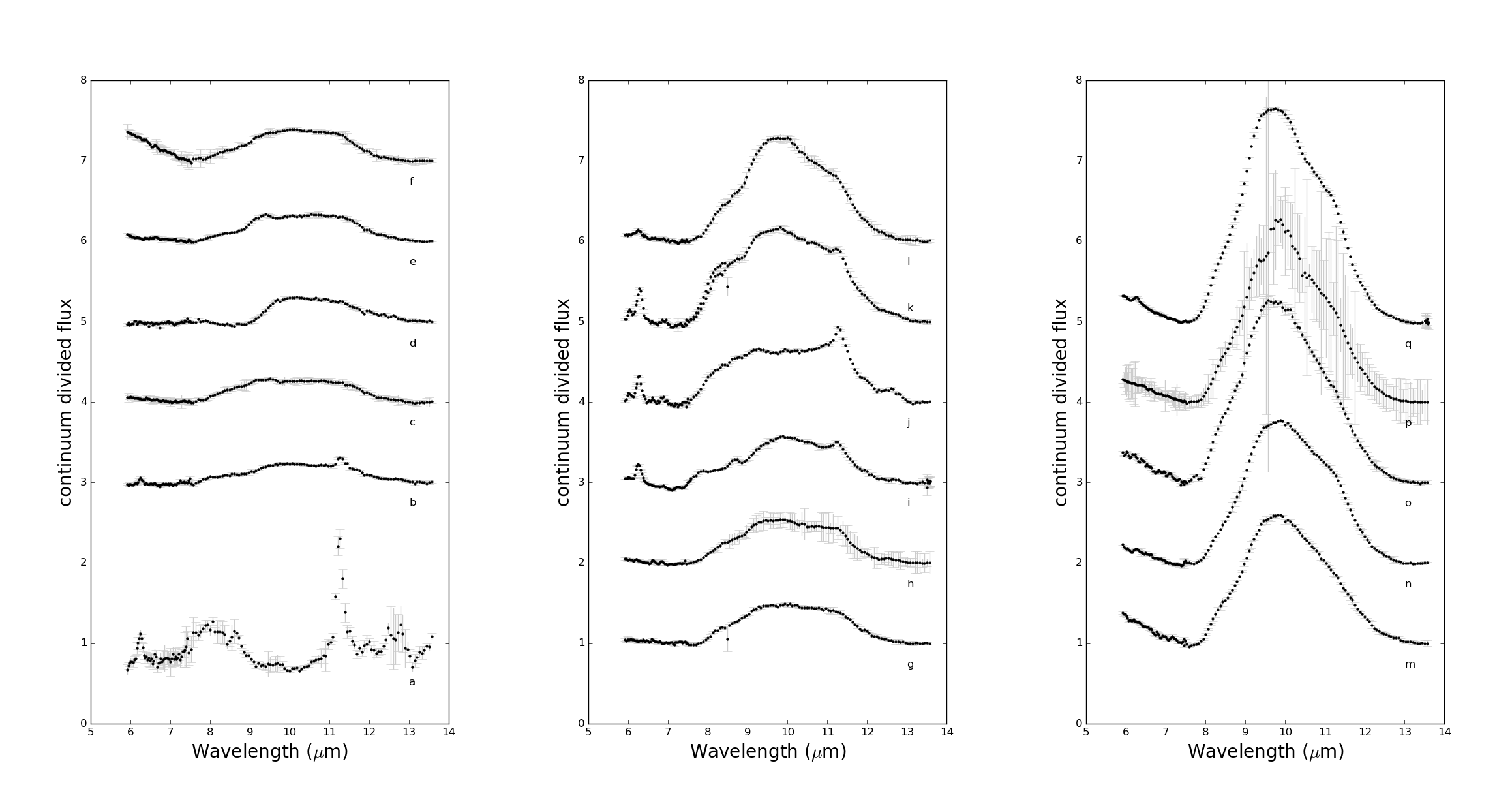}
        \caption{The continuum divided normalized peak strength over the 10 $\mu$m feature for those sources classified as Group II. The continuum between 7.5 and 13.5 $\mu$m has been normalized to 1. Only a) V2149 Ori, do not have a detectable silicate feature. The other sources then follow in ascending order of peak strength: b) HD 294260, c) CO Ori, d) PR Ori, e) HD 35929, f) DI Cha, g) BE Ori, h) HT Lup, i) HD 142666, j) GX Ori, k) HBC 442 l) V1650 Ori, m) CR Cha, n) CV Cha, o) RY Ori, p) RY Tau, q) HD 144432. The scale is indicated in the leftmost panel and is the same for all sources, spectra are shifted to allow comparison.}
        \label{fig:C-GIIsil}
    \end{figure*}

    To measure the strength and shape of the 10 ${\mu}m$ silicate feature we first measured the continuum flux at 7 ${\mu}m$ and at 13.5 ${\mu}m$ where no prominent emission or absorption bands are present. The underlying continuum was then approximated to be linear between these two points and used to normalize the spectra. We have defined the strength of the 10${\mu}m$ band by measuring the continuum divided flux at 9.8 ${\mu}m$ ($F_{9.8}$) and the shape by the ratio of the continuum divided flux at 11.3${\mu}m$ and at 9.8 ${\mu}m$ (F$_{11.3}$/F$_{9.8}$). We set the threshold for a detectable silicate feature to a peak over continuum strength of 1.2 to clearly separate the emission from the noise in the spectra. In Figure \ref{fig:C-grainevol} we show the relationship between shape of the 10 $\mu$m silicate feature and the peak strength at 9.8 $\mu$m. In figure \ref{fig:C-GIsil} and \ref{fig:C-GIIsil} we show the shape of the 10 $\mu$m silicate feature as a function of rising peak strength for Group I and Group II sources respectively. The 3 sources that are in the category Group I/II are shown together with the group of which their spectral index, $[F_{30}/F_{13.5}]$, lies closest. The absence of a 10 $\mu$m silicate feature means that there are no small grains of the correct temperature for the emission to be formed. This can come from either a gap in the disk where no silicate grains are present, or that the small grains have grown to a size where the emission is not longer present.

    Following the HAeBe star classification \citep{2001AA...365..476M}, we call the disks with silicate feature present Ia and IIa and the disks showing no silicate feature Ib and IIb. Using the Spitzer spectra we have available we are able to make this sub categorization for 15 Group I disks and 17 Group II disks. We find 9 Group Ia disks vs. 6 Group Ib disks and 17 Group IIa disks and no group IIb disk. The 3 disks with spectral index that puts them in between Group I and Group II, 2 are Group Ia/IIa disks and one Group Ib/IIb disk. Among the HAeBe stars the distribution is similar with 15 Group Ia vs. 5 Group Ib and 29 Group IIa vs. 4 Group IIb \citep{2010ApJ...721..431J}.

\begin{figure}
        \includegraphics[width=8.8cm]{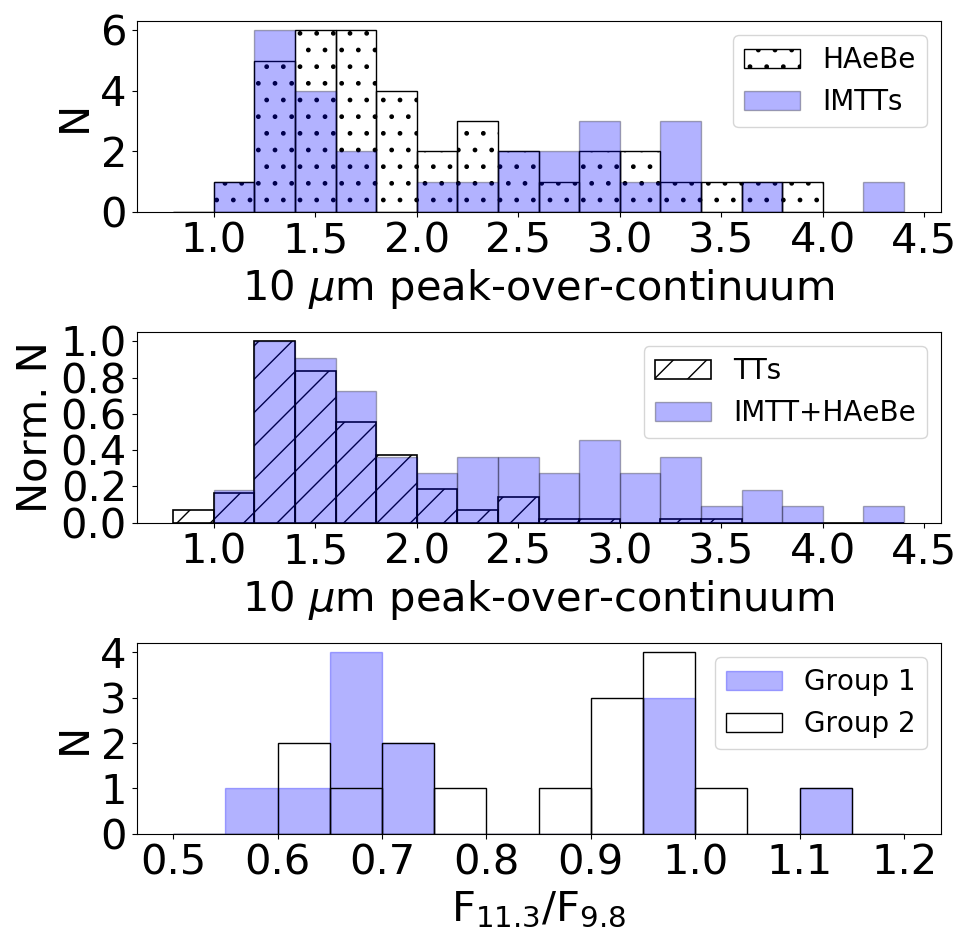}
        \caption{\textbf{Upper panel:} The 10 $\mu$m peak-over-continuum distribution among the HAeBe stars \citet{2010ApJ...721..431J} in comparison with the IMTT star sample. \textbf{Middle panel:} The 10 $\mu$m peak-over-continuum distribution among T-Tauri stars \citep{2011ApJS..195....3F} in comparison with the intermediate mass star samples (IMTTs+HAeBe). Because of the large difference in sample size the distributions have been normalized so that their peak value equals 1. \textbf{Lower panel:} The distribution in the F$_{11.3}$/F$_{9.8}$ ratio clearly showing the separation between the large grains group and the smaller grains group among the IMTT stars and that Group 1 disks more commonly have smaller grains.}
        \label{fig:C-Sicomp}
    \end{figure}

   We find the relationship between the flux ratio [$F_{11.3}/F_{9.8}$] and the peak over continuum flux at 9.8 $\mu$m, in Group Ia and Group IIa disks, follows the same relationship as found among HAeBe and T-Tauri stars by other authors \citep{2018AA...617A..83V, 2003AA...400L..21V,2008ApJ...683..479B, 2003AA...412L..43P,2009ApJ...703.1964F}. However, the peak-over-continuum strength distribution of the 10 $\mu$m silicate feature is more similar to the HAeBe stars than the T-Tauri star distribution (see figure \ref{fig:C-Sicomp}), where the T-Tauri stars lack the tail towards higher emission band strength visible in both the IMTT stars and HAeBe stars. Such a trend extends over a much wider stellar and sub-stellar mass range \citep[e.g.][]{2009ApJ...696..143P}, and is interpreted in terms of an increase in the degree of inner disk settling with decreasing stellar mass, extending to the brown dwarf regime.

    We also note that the distribution of disks in the [$F_{11.3}/F_{9.8}$] flux ratio clusters around two values: one with low peak over continuum intensity at 9.8 $\mu$m (weak feature) and with [$F_{11.3}/F_{9.8}$] flux ratio of around 0.9-1.1, consistent with larger grains (around 2 $\mu$m), and one region with high  peak over continuum intensity (strong feature) and low [$F_{11.3}/F_{9.8}$] flux ratio (0.6-0.7) consistent with smaller grains (around 0.1 $\mu$m).  This could be related to changes in the characteristic grain size from sub-micron to a few microns, and the corresponding non-linear change in grain opacity.
    
     \citet{2013AA...555A..64M} show that HAeBe stars with weak or no silicate emission have a [$F_{30}/F_{13.5}$] $\gtrsim$ 5 \citep[see also][]{2018AA...617A..83V}.  These disks have large cavities, and a corresponding lack of small warm silicate grains that are responsible for the 10 $\mu$m silicate band.  This aspect is also reflected in  the IMTT star sample (table \ref{tbl:Flux_mesurements}). 
     The majority of IMTT stars with strong silicate features have [$F_{30}/F_{13.5}$] flux ratios well below 5. We can therefore conclude that the emission from silicate grains around 10 $\mu$m behaves like in the HAeBe stars, where the small warm silicate grains are much less abundant in the inner disk due to large cavities. 
     
     If  the hot innermost disk contains small silicate grains, it should still produce a weak silicate band. In two cases, HD~135344B \citep{2018AA...617A..83V} and the lower mass T Tauri star T~Cha  \citep[which hosts a transitional disk, see][]{2013AA...552A...4O}, interferometric observations suggest that such small hot silicate grains are absent. Therefore the lack of silicate emission in some group I disks is likely due to a combination of a large disk gap and a depletion of small grains in the inner disk.

\subsection{Polycyclic Aromatic Hydrocarbons (PAHs)}

    We use low resolution Spitzer spectra between 6 and 13 $\mu$m to detect the presence of any PAH emission. The PAH emission bands in this wavelength range are at 6.2 $\mu$m, 7.7 $\mu$m, 8.6 $\mu$m, 11.3 $\mu$m and 12.7 $\mu$m. We inspect the Spitzer spectra visually and if we see 2 clear PAH bands in the spectra we count this as a detection. In the cases we can only see one clear band we define the PAH detection to be tentative. The result is presented in table \ref{tbl:PAH_detect}. The spectral features at 7.7 $\mu$m and 8.6 $\mu$m sometimes blend into the blue edge of 10 $\mu$m silicate features and are therefore often hard to distinguish. The same situation happens on the red edge of the 10 $\mu$m silicate feature with the 11.3 $\mu$m PAH band that can even disappear with strong amorphous silicate emission.
    
    We find that the PAH detection rate among the IMTT stars is 27\%. Considering also tentative detections the detection rate becomes 44\%. In comparison, the detection rate of PAH emission among HAeBe stars is 70\% \citep{2010ApJ...718..558A} and only ~8\% among T-Tauri stars \citep{2006AA...459..545G}. This positions the IMTT stars, with a higher average luminosity and temperature than the low mass T-Tauri stars but a weaker UV field than the HAeBe stars, in between these two groups in terms of detection rate. The infrared emission from PAH-molecules, for isolated stars, is caused by stellar UV photons that excite electronic states in the PAH molecule, which subsequently de-excites through vibrational emission in stretching and bending mode resonances. The PAH detection rate correlates with the $T_{\rm eff}$ of the star since the UV radiation field is stronger the hotter the star is \citep{2006ApJS..165..568F,2010ApJ...718..558A}. 
    
    In order for PAHs to be excited they need be in a region optically thin to UV radiation. This could be in the surface of a flaring disk \citep[ex. HD 97048,][]{2006Sci...314..621L}, the inner rim of the disk or in a region where small dust grains are removed while PAHs are still present \citep[e.g. disk gaps or in settled disks,][]{2007AA...469L..35G,2016AA...586A.103W}. It is therefore interesting to consider the PAH detection rate and strength as a function of disk geometry. In the HAeBe sample the PAH detection rates for group Group I and Group II sources are 80\% and 64\%, respectively, and PAH emission is generally stronger in Group I disks. Both observations can be understood because Group I sources show large gaps and/or dust depleted inner disk regions where the UV photons can freely propagate. In contrast, in group II disks the PAH emission is likely to come from a decoupling of the gas from the dust where the dust has settled towards the mid-plane but the gas is still flaring \citep{2010ApJ...718..558A}; or the disk is small.  
    
    We find that the PAH detection rate in the IMTT Group I and Group II sources are 47\% and 17\% respectively. Adding the tentative rate it becomes 73\% Group I and 29\% Group II. We find that the Group I disks around IMTT stars have on average stronger PAH bands than in Group II disks, as is the case for the HAeBe stars. We conclude that the main difference between HAeBe stars and IMTT is the smaller strength of the PAH bands in the IMTT spectra, but that otherwise both samples show more frequent detection in Group I disks.
    
    Some the most prominent PAH features are seen in SR 21, Haro 1-6 and V2149 Ori. The PAH emission around V2149 Ori, which is classified as a Group I/II disk, is likely to arise from the disk \citep{2013ApJ...769..149K} but due to its location, 9 arcmin away from M42s center just outside the HII region, the radiation exiting the PAH might come from the surrounding hotter stars and not from the binary itself.

    The intensity of UV radiation also affects the chemistry of the PAH molecules \citep{2010ApJ...718..558A}. In our sample, we find that the peak of the 6.2 and 7.7 micron PAH bands are often shifted to 6.3 and 7.8 $\mu$m. One explanation for this could be that stronger UV radiation field increases the ratio of aromatic to aliphatic hydrocarbons \citep{2007ApJ...664.1144S,2008ApJ...684..411K}. When the UV radiation field decreases the 6.2 and 7.7 $\mu$m emission band shifts towards the red. It can be observed that when $T_{\rm eff}$ decreases these emission bands experience this shift consistent with lower UV flux from the central star.  The emission at 6.2 and 7.7 $\mu$m emission is red-shifted in many of the spectra something that can be seen also in cooler T Tauri stars \citep{2010ApJ...718..558A}.

\section{Discussion}

    We have constructed a sample of nearby stars, the IMTT stars, that we consider to be likely progenitors of the class of well studied HAeBe stars.  We did so in order to facilitate a better understanding of the evolution of gas-rich protoplanetary disks. We here discuss our findings and try to give them an interpretation. 

\subsection{Sample and stellar parameters}
    The masses of the IMTT stars found using our method of selection are predominately located in the lower region of the selection mass interval. The median mass of the sample is 1.87$M_{\odot}$. This is partly because of the IMF but mostly because we have created an optically selected sample of the most massive T-Tauri stars. The method only includes stars that have already cleared their stellar envelope which is confirmed by that most stars in the sample show a low to moderate reddening (see table \ref{tbl:Stellarparam}). This means we do not have the very youngest and more massive stars in our sample. The lack of higher mass stars ($\geq 4 M_{\odot}$) in our specified spectral type range is consistent with a shorter evolutionary timescale for higher mass stars, and the fact that they are less frequent.  Within our distance limit of 500~pc, we expect very young and massive PMS stars to be present only in the Orion SFR. For stars having a high extinction (for example a high inclination, near edge-on disk or remaining cloud material in a Lada class 0/I object) our luminosity and mass estimates could be lower than the real values.
    
    The \textit{Gaia} DR2 parallaxes for the final sample where compared with the new EDR3 release parallaxes. No significant differences where found for stars with RUWE$\leq$ 1.4.

    To determine the mass and age we use \citet{2000AA...358..593S} pre-main sequence stellar evolutionary tracks using a solar like metallicity (z=0.01). We note that stellar evolution is dependent on metallicity and that in turn masses and ages are affected by the choice of metallicity in the stellar models. A higher metallicity results in higher masses and older stars. However assuming a solar metallicity in the solar neighbourhood corresponds well with G, F and early B-type stars \citep{2001ApJ...554L.221S,2008ApJ...688L.103P}.

\subsection{The disks}
    We have shown that we find similar disks to the HAeBe sample of \citet{2010ApJ...721..431J} where the Group I disks are 38\% of the population. Among the IMTT stars there is a more even distribution with 51\% Group I which is not too different from the HAeBe stars given the sample size. Comparing the SEDs we see that they are similar except for the most extreme 7$\mu m$ excess cases that can be understood by bolometric correction effects. Comparing the spatially resolved data in literature for the IMTT star sample with that of the HAeBe stars we see that the disks around IMTT stars also show gaps, rings and spirals suggesting disks around IMTT stars are not so very different from those around the HAeBe stars.
    
    60\% of the Group I disks and all Group II disks show silicate emission at 10$\mu$m in our sample. In the HAeBe sample \citep{2010ApJ...721..431J} silicate emission is present in 75\% of the Group I disks and 87\% of the Group II disks. The relationship between the strength and shape of the 10 $\mu$m silicate feature (figure \ref{fig:C-grainevol} and figure \ref{fig:C-Sicomp} top and middle panel) is consistent with the relationship found among the HAeBe stars \citep{2005AA...437..189V} which suggests that in terms of silicate grain size and growth the grains in IMTT disks are very similar to the HAeBe disks. 
    The emission from silicate dust grains at 10 $\mu$m in the IMTT sample can be seen in more than half the Group I disks, with a wide silicate strength distribution, from PAH dominated disk without a silicate feature to very strongly peaked features. A silicate emission feature at 10 $\mu$m is present in all Group II disks (with the exception of V2149 Ori) with on average lower peak strength than in the Group I sources. This may be explained in the following way: The small silicate grains that gives rise to the solid state feature at 10 $\mu m$ are missing in some Group I disks, either because they have grown to a distribution above $\sim$5 $\mu m$ \citep{2013AA...552A...4O} or because the disk has a cleared-out cavity at the desired temperature and location where the emission is normally coming from \citep{2013AA...555A..64M}. In shape and strength the IMTT 
    silicate feature cluster in two groups: one with small pristine grains (strong, peaked feature) and one consistent with more processed grains (weak, broad feature). This can be seen in figure \ref{fig:C-Sicomp}. This grouping is not present in the silicate emission feature distribution for the T-Tauri stars \citep{2009ApJ...703.1964F} nor for the HAeBe stars \citep{2010ApJ...721..431J}. About twice as many Group I sources show signs of more pristine small silicate grains while for the Group II disks seems to the opposite with more disks having a more processed grain population with larger silicate grains.

    
    
    
    The detection rate for emission of PAH-molecules in the sample is 27\%. This puts the IMTT stars in between the HAeBe (about 70\%) and the T-Tauri stars (about 8\%). The PAH bands are generally weak with only a few exceptions. This is most likely because IMTT stars are cooler than the HAeBe stars and therefore have a lower UV flux, needed to excite the PAHs. As luminosity and temperature decreases so does the relative contrast of the PAH emission to the continuum and the PAHs disappear among the classical T-Tauri stars. The relatively weak PAH emission could also be a reason for why we find weaker silicate emission in our sample and why the Group Ib disks are less frequent in the IMTT star sample than among the HAeBe stars, since weaker PAH emission would make any low level silicate emission easier to distinguish. 

    We find that spatially resolved data available in the literature (see Appendix C) confirms our classification in most cases. Disk features such as spirals and gaps are also present among the IMTT stars suggesting that the disks are already evolved. There are indications that planets form at early stages of disk dissipation and that the process happens quickly \citep{2018ApJ...869L..41A,2018ApJ...869...17L,2018ApJ...869L..47Z,2019ApJ...872..112V} and the transition disk lifetime is thought to be short, $\sim 0.5$ Myr \citep{2007ApJ...667..308C}. Looking at the similarity between IMTT and HAeBe disks this suggests that disk dissipation and planet formation starts early in some disks and later in others. Considering that disk lifetimes are typically of the order of a few Myr \citep{2014prpl.conf..475A, 2007ApJ...667..308C}, it is interesting that the disks around IMTT stars are so similar to the HAeBe stars given that the timescale over which the IMTT star evolves into a HAeBe star is considerably longer.

    If there was a evolutionary link between the pre-transitional Group I and Group II disks, one would expect the Group II disks to be the dominant disk type at this earlier stage in the evolution since disk dispersal happens from the inside out in the disk \citep{2013MNRAS.428.3327K}, something that is not apparent from the sample presented here. For these disks to show up in both the HAeBe and IMTT star sample they need to have a significant lifetime.
    
    Disk gaps can form early in the IMTT stars as well as late in the HAeBe stars or maybe even earlier and persist for a long time period. In order to prevent the inner disk in a gapped disk from quickly empty onto the star, it needs to be continuously replenished by the outer disk gas, and some dust, that crosses the gap to the inner disk. If the disk dissipation can be postponed to the later stages of HAeBe stars, one may expect to find a higher fraction of IMTT stars with gapless disks. But our sample contains roughly equal amounts of Group I (gapped) and Group II (self-shadowed) disks. In addition, the ratio is not very different from that in the HAeBe stars. This may point to a lifetime of gapped disks that is significant. It is hard to imagine a scenario where small Group II disks evolve into large Group I disks. It suggests that Group I and Group II disk evolution are disconnected from each other. Maybe Group I and Group II disks evolve from different parent populations whose parameters are set by the environment at a very early stage.

\section{Conclusions}
    The constructed list will serve as a basis for further studies of the IMTT stars and their disks. The sample is not a complete definite list and is likely to grow as more of these stars are identified. 

   \begin{enumerate}
      \item With our selection method and criteria we find 49 IMTT stars with infrared excess out of which 6 are debris disks and 44 gas rich disks. Some have been studied earlier as lower mass HAeBe stars and 10 are new IMTT stars previously classified as T-Tauri stars.
      \item The sample disks show a more even distribution between Group I and Group II disk geometries than HAeBe samples. 
      \item The frequency of detected silicate emission is about the same as in the HAeBe samples per disk group and the relationship between strength and shape of the 10 $\mu$m emission follow the same anti-correlation as the HAeBe stars and points towards grain sizes that are comparable to those of the HAeBe stars.
      \item The reddest disks we find [$F_{30}/F_{13.5}$] $\geq 5$ are consistent with Group I disks with large central cavities and weak or no silicate emission.
      \item The presence of PAH emission is less frequent in the IMTT stars spectra in the sample than among the HAeBe stars but more frequent than among the T-Tauri stars. This is probably caused by an on average decreasing effective temperature of HAeBe stars, IMTT stars and lower mass T Tauri stars, respectively, and the corresponding lower UV flux.
      \item Disk dissipation (and therefore planetary formation) seems to takes place early in some disks and not so early in others.
      \item Implications are that Group I and Group II disks are two different evolutionary paths a disk can take; and that this is determined at a very early stage of evolution. This is consistent with the conclusion drawn by \cite{2017AA...603A..21G}.
   \end{enumerate}

\begin{acknowledgements}
      This publication makes use of data products from the Wide-field Infrared Survey Explorer, which is a joint project of the University of California, Los Angeles, and the Jet Propulsion Laboratory/California Institute of Technology, funded by the National Aeronautics and Space Administration. 
      This work has made use of data from the European Space Agency (ESA) mission {\it Gaia} (\url{https://www.cosmos.esa.int/Gaia}), processed by the {\it Gaia} Data Processing and Analysis Consortium (DPAC, \url{https://www.cosmos.esa.int/web/Gaia/dpac/consortium}). Funding for the DPAC has been provided by national institutions, in particular the institutions participating in the {\it Gaia} Multilateral Agreement. 
      This research has made use of the SIMBAD database, operated at CDS, Strasbourg, France. This research has made use of the VizieR catalogue access tool, CDS, Strasbourg, France (DOI: 10.26093/cds/vizier). The original description of the VizieR service was published in A\&AS 143, 23
      We would also like to thank MSc. Erik Gisolf for his contribution to this paper. Finally we thank the anonymous referee for constructive comments that improved the paper significantly.

\end{acknowledgements}

    \begin{figure*}[ht]
        \label{fig:C-SEDgal}
        \caption{The SED with fitted Kurucz model. The Kurucz model is in blue, literature photometry is black (see references table \ref{tbl:phot_ref}) and Spitzer spectra in red.}
        \includegraphics[width=\textwidth]{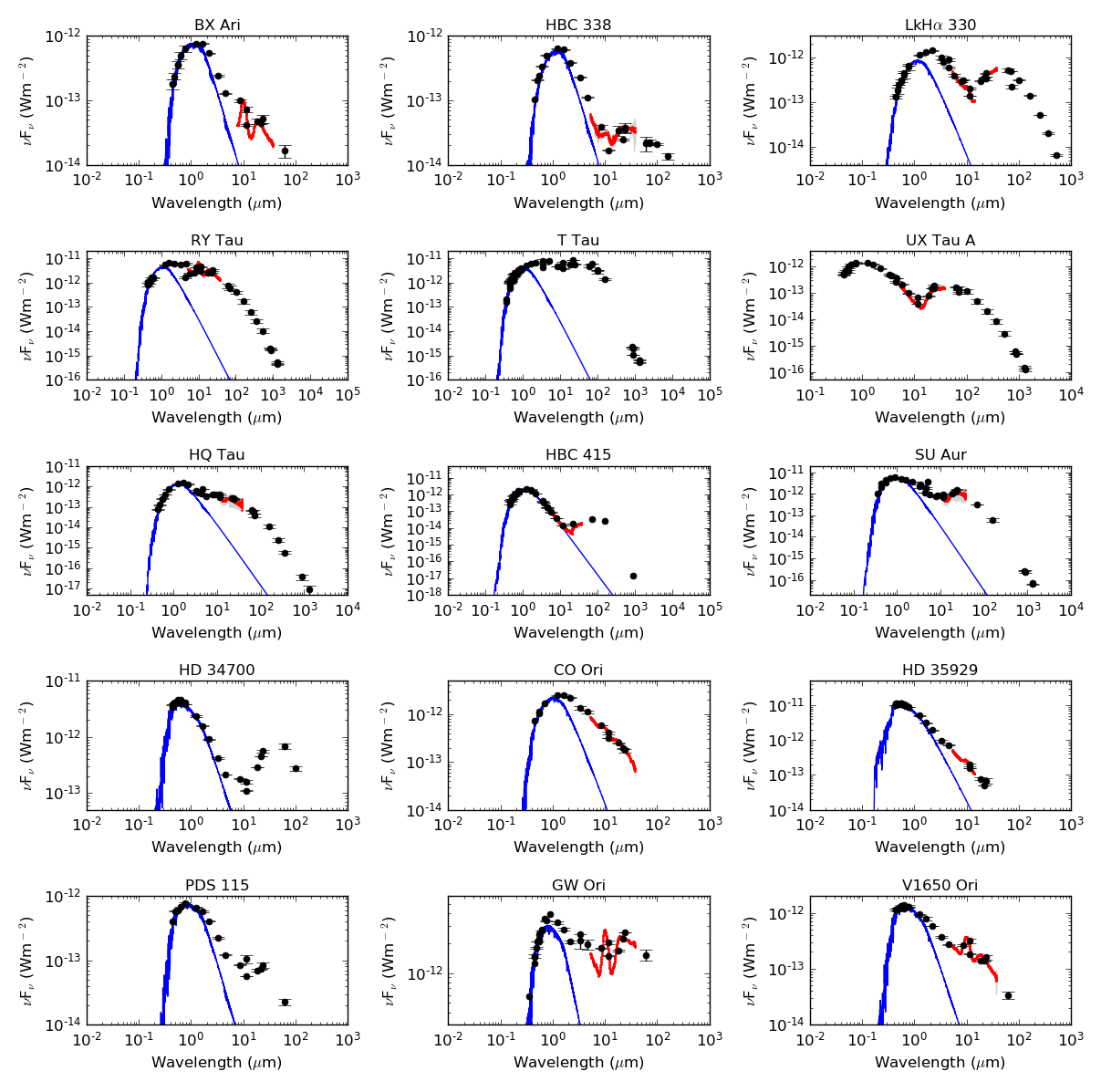}
    \end{figure*}

    \begin{figure*}[ht]
        \label{fig:C-SEDgal2}
            \includegraphics[width=\textwidth]{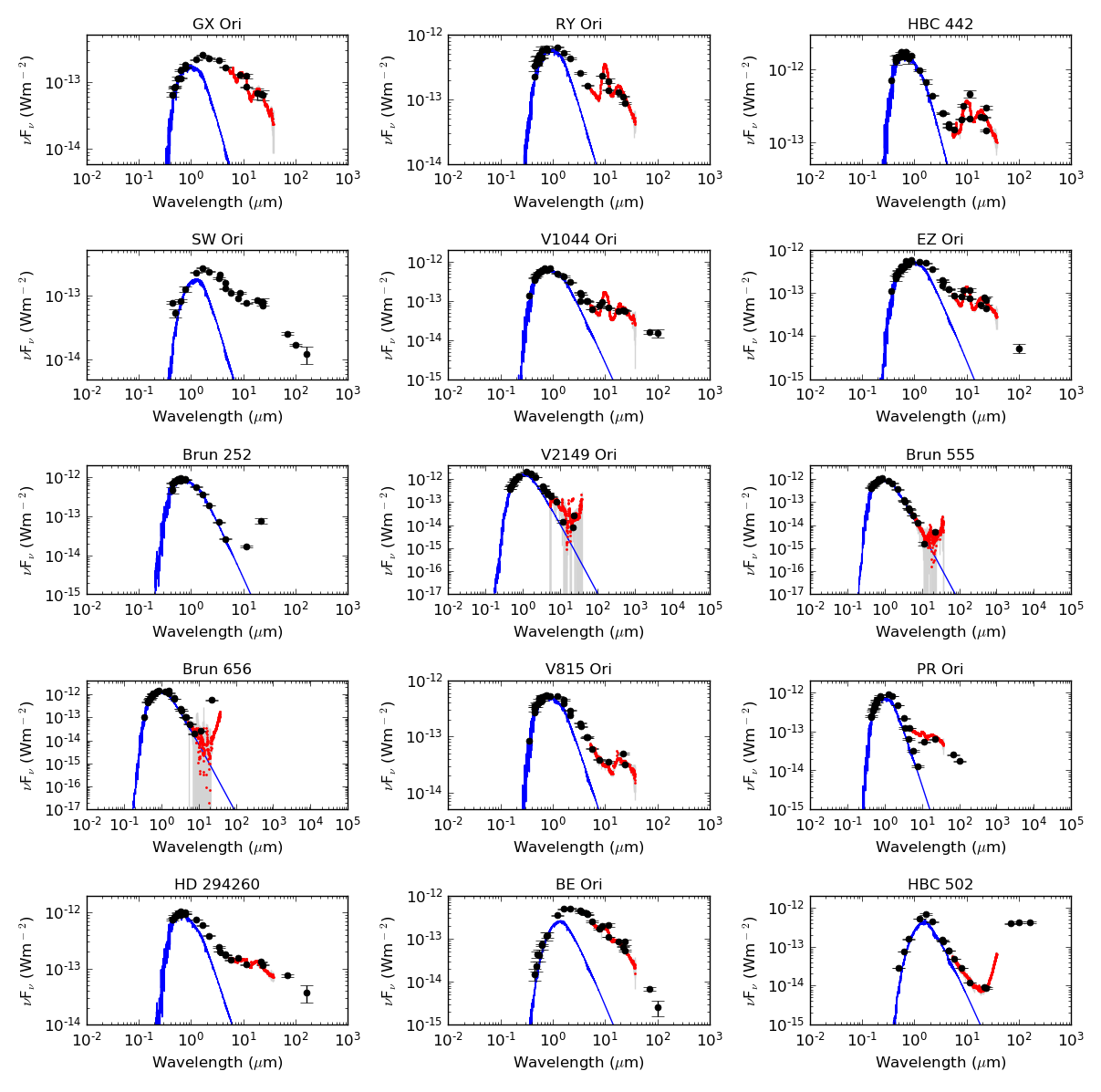}
    \end{figure*}
    \begin{figure*}[ht]
        \label{fig:C-SEDgal3}
            \includegraphics[width=\textwidth]{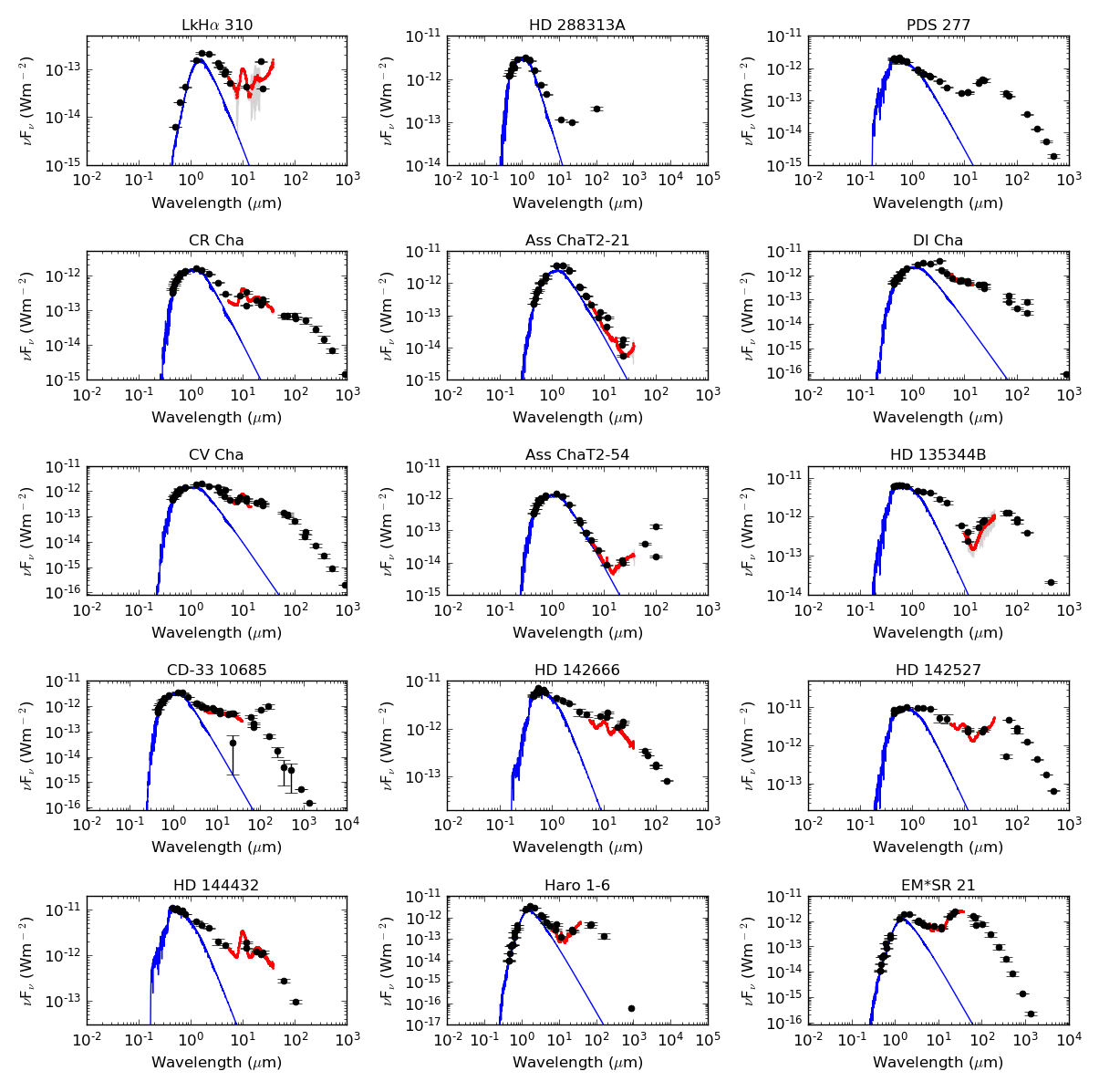}
    \end{figure*}
    \begin{figure*}[ht]
        \label{fig:C-SEDgal4}
            \includegraphics[width=\textwidth]{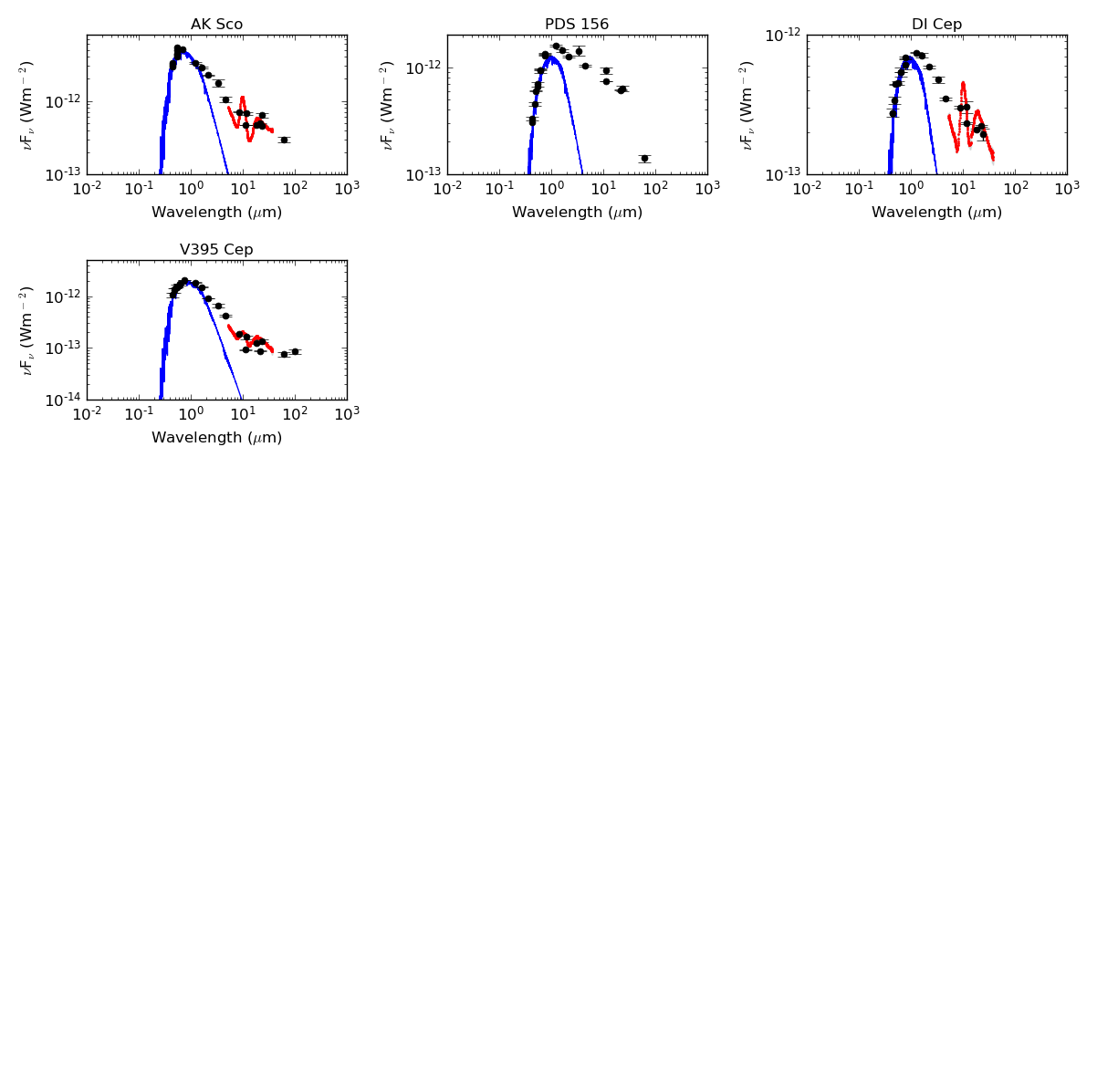}
    \end{figure*}


\begin{sidewaystable*}
\caption{Fluxes as measured from Spitzer spectra used in section 3 together with calculated flux ratios presented. }
\label{tbl:Flux_mesurements}
\scriptsize
\begin{tabular}{lrrlllllllllllr}
\\
\hline \hline
            & \multicolumn{2}{c}{Kurucz Model} & \multicolumn{6}{l}{Measurements from Spectra and Photometry} & \multicolumn{3}{l}{Computes Excesses} & \multicolumn{3}{l}{Computed Flux Ratios}                                                                                                                                                                           \\
Star         & F7 [Jy] & F60 [Jy] & F7 [Jy]           & F9.8 [Jy]          & F11.3 [Jy]         & F13.5 [Jy]        & F30 [Jy]           & F60 [Jy]            & E[7] [mag]      & E[60] [mag]      &  & $\frac{F30}{F13.5}$  & $\frac{F13.5}{7}$ & $\frac{FN11.3}{FN9.8}$ \\
\hline
BX Ari       & 0.0175  & 0.0005   &                   & \(0.337\pm0.003\)  & \(0.234\pm0.003\)  & \(0.119\pm0.002\) & \(0.261\pm0.004\)  & \(0.342\pm0.075\)   &                 & \(7.05\pm0.22\)  &  & \(2.193\pm0.050\)    &                   & 0.671                  \\
HBC 338      & 0.0299  & 0.0005   & \(0.076\pm0.004\) & \(0.102\pm0.003\)  & \(0.118\pm0.003\)  & \(0.094\pm0.006\) & \(0.367\pm0.014\)  & \(0.442\pm0.111\)   & \(1.01\pm0.06\) & \(7.48\pm0.24\)  &  & \(3.904\pm0.290\)    & \(1.237\pm0.083\) & 1.100                  \\
LkH$\alpha$ 330     & 0.0426  & 0.0007   & \(0.663\pm0.005\) & \(0.749\pm0.006\)  & \(0.669\pm0.005\)  & \(0.455\pm0.008\) & \(4.864\pm0.028\)  & \(10.700\pm1.070\)  & \(2.98\pm0.01\) & \(10.54\pm0.10\) &  & \(10.690\pm0.198\)   & \(0.686\pm0.019\) & 0.952                  \\
RY Tau       & 0.2013  & 0.0031   & \(5.753\pm0.036\) & \(22.857\pm0.248\) & \(16.323\pm0.378\) & \(9.003\pm0.127\) & \(17.721\pm0.096\) & \(15.300\pm3.060\)  & \(3.64\pm0.01\) & \(9.25\pm0.20\)  &  & \(1.968\pm0.030\)    & \(1.565\pm0.015\) & 0.643                  \\
T Tau        & 0.1608  & 0.0024   &                   &                    &                    &                   &                    & \(98.700\pm11.844\) &                 & \(11.52\pm0.12\) &  &                      &                   &                        \\
UX Tau A     & 0.0857  &          & \(0.238\pm0.006\) & \(0.150\pm0.004\)  & \(0.153\pm0.004\)  & \(0.120\pm0.007\) & \(1.453\pm0.047\)  & \(3.480\pm0.174\)   & \(1.11\pm0.03\) &                  &  & \(12.108\pm0.808\)   & \(0.504\pm0.064\) & 1.142                  \\
HQ Tau       & 0.0854  & 0.0013   &                   &                    & \(1.384\pm0.004\)  & \(0.931\pm0.003\) & \(1.480\pm0.087\)  & \(1.390\pm0.125\)   &                 & \(7.56\pm0.09\)  &  & \(1.590\pm0.094\)    &                   &                        \\
HBC 415      & 0.0984  & 0.0015   & \(0.120\pm0.003\) & \(0.062\pm0.002\)  & \(0.050\pm0.002\)  & \(0.041\pm0.005\) & \(0.161\pm0.013\)  &                     & \(0.22\pm0.03\) &                  &  & \(3.927\pm0.574\)    & \(0.342\pm0.124\) & 1.028                  \\
SU Aur       & 0.1993  & 0.0030   &                   &                    & \(2.996\pm0.008\)  & \(2.089\pm0.024\) & \(9.841\pm0.530\)  &                     &                 &                  &  & \(4.711\pm0.055\)    &                   &                        \\
HD 34700     & 0.0662  & 0.0010   &                   &                    &                    &                   &                    & \(14.100\pm1.410\)  &                 & \(10.43\pm0.10\) &  &                      &                   &                        \\
CO Ori       & 0.0870  & 0.0013   & \(1.386\pm0.015\) & \(1.712\pm0.025\)  & \(1.685\pm0.016\)  & \(1.353\pm0.020\) & \(1.191\pm0.035\)  &                     & \(3.01\pm0.01\) &                  &  & \(0.880\pm0.029\)    & \(0.976\pm0.018\) & 0.980                  \\
HD 35929     & 0.1168  & 0.0017   & \(0.740\pm0.009\) & \(0.827\pm0.008\)  & \(0.766\pm0.007\)  & \(0.545\pm0.012\) &                    &                     & \(2.00\pm0.01\) &                  &  & \(0.000\pm\)         & \(0.736\pm0.025\) & 0.998                  \\
PDS 115      & 0.0230  & 0.0003   &                   &                    &                    &                   &                    & \(0.471\pm0.052\)   &                 & \(7.86\pm0.11\)  &  &                      &                   &                        \\
GW Ori       & 0.0869  & 0.0013   & \(2.531\pm0.035\) & \(0.923\pm0.039\)  & \(7.134\pm0.031\)  & \(4.323\pm0.029\) & \(20.006\pm0.138\) & \(31.500\pm4.095\)  & \(3.66\pm0.01\) & \(10.98\pm0.13\) &  & \(4.628\pm0.045\)    & \(1.708\pm0.015\) & 0.570                  \\
V1650 Ori    & 0.0252  & 0.0004   & \(0.457\pm0.009\) & \(1.217\pm0.028\)  & \(1.008\pm0.020\)  & \(0.634\pm0.035\) & \(0.933\pm0.023\)  & \(0.688\pm0.103\)   & \(3.14\pm0.02\) & \(8.19\pm0.15\)  &  & \(1.472\pm0.089\)    & \(1.387\pm0.059\) & 0.764                  \\
GX Ori       & 0.0068  & 0.0001   & \(0.258\pm0.006\) & \(0.415\pm0.007\)  & \(0.473\pm0.006\)  & \(0.253\pm0.007\) & \(0.360\pm0.015\)  &                     & \(3.96\pm0.02\) &                  &  & \(1.423\pm0.071\)    & \(0.981\pm0.036\) & 1.121                  \\
RY Ori       & 0.0208  & 0.0003   & \(0.255\pm0.006\) & \(1.146\pm0.019\)  & \(0.878\pm0.030\)  & \(0.503\pm0.012\) & \(0.693\pm0.027\)  &                     & \(2.72\pm0.03\) &                  &  & \(1.378\pm0.063\)    & \(1.973\pm0.034\) & 0.648                  \\
HBC 442      & 0.0281  & 0.0004   & \(0.325\pm0.006\) & \(1.218\pm0.019\)  & \(1.278\pm0.013\)  & \(0.849\pm0.013\) & \(1.400\pm0.078\)  &                     & \(2.66\pm0.02\) &                  &  & \(1.649\pm0.095\)    & \(2.612\pm0.024\) & 0.860                  \\
SW Ori       & 0.0097  & 0.0001   &                   &                    &                    &                   &                    &                     &                 &                  &  &                      &                   &                        \\
V1044 Ori    & 0.0176  & 0.0003   & \(0.114\pm0.005\) & \(0.559\pm0.009\)  & \(0.424\pm0.009\)  & \(0.148\pm0.006\) & \(0.382\pm0.017\)  &                     & \(2.03\pm0.05\) &                  &  & \(2.581\pm0.155\)    & \(1.298\pm0.060\) & 0.710                  \\
EZ Ori       & 0.0172  & 0.0003   & \(0.132\pm0.006\) & \(0.467\pm0.006\)  & \(0.370\pm0.006\)  & \(0.188\pm0.005\) & \(0.418\pm0.020\)  &                     & \(2.21\pm0.05\) &                  &  & \(2.223\pm0.122\)    & \(1.424\pm0.053\) & 0.721                  \\
Brun 252     & 0.0173  & 0.0003   &                   &                    &                    &                   &                    &                     &                 &                  &  &                      &                   &                        \\
V2149 Ori    & 0.0699  & 0.0011   & \(0.252\pm0.051\) & \(0.180\pm0.037\)  & \(0.419\pm0.057\)  & \(0.160\pm0.062\) & \(0.333\pm0.132\)  &                     & \(1.39\pm0.20\) &                  &  & \(2.081\pm1.154\)    & \(0.635\pm0.437\) & 2.438                  \\
Brun 555     & 0.0313  & 0.0005   & \(0.034\pm0.003\) & \(0.018\pm0.003\)  & \(0.011\pm0.006\)  & \(0.011\pm0.005\) & \(0.059\pm0.013\)  &                     & \(0.09\pm0.09\) &                  &  & \(5.364\pm2.709\)    & \(0.324\pm0.463\) & 0.687                  \\
Brun 656     & 0.0464  & 0.0007   & \(0.077\pm0.018\) & \(0.051\pm0.069\)  & \(0.068\pm0.019\)  & \(0.015\pm0.104\) & \(0.408\pm0.024\)  &                     & \(0.55\pm0.23\) &                  &  & \(27.200\pm188.593\) & \(0.195\pm6.937\) & 1.622                  \\
V815 Ori     & 0.0148  & 0.0002   & \(0.100\pm0.005\) & \(0.100\pm0.003\)  & \(0.135\pm0.005\)  & \(0.097\pm0.005\) & \(0.292\pm0.028\)  &                     & \(2.08\pm0.05\) &                  &  & \(3.010\pm0.328\)    & \(0.970\pm0.072\) & 1.322                  \\
PR Ori       & 0.0279  & 0.0004   & \(0.216\pm0.006\) & \(0.319\pm0.012\)  & \(0.334\pm0.007\)  & \(0.291\pm0.010\) & \(0.593\pm0.018\)  &                     & \(2.22\pm0.03\) &                  &  & \(2.038\pm0.093\)    & \(1.347\pm0.044\) & 0.973                  \\
HD 294260    & 0.0159  & 0.0002   & \(0.318\pm0.062\) & \(0.485\pm0.009\)  & \(0.552\pm0.007\)  & \(0.476\pm0.010\) & \(0.839\pm0.028\)  &                     & \(3.25\pm0.19\) &                  &  & \(1.763\pm0.070\)    & \(1.497\pm0.196\) & 1.030                  \\
BE Ori       & 0.0183  & 0.0003   & \(0.421\pm0.007\) & \(0.604\pm0.012\)  & \(0.563\pm0.013\)  & \(0.404\pm0.009\) & \(0.356\pm0.012\)  &                     & \(3.40\pm0.02\) &                  &  & \(0.881\pm0.036\)    & \(0.960\pm0.028\) & 0.934                  \\
HBC 502      & 0.0369  & 0.0006   & \(0.075\pm0.004\) & \(0.059\pm0.003\)  & \(0.054\pm0.004\)  & \(0.044\pm0.004\) & \(0.210\pm0.004\)  &                     & \(0.77\pm0.06\) &                  &  & \(4.773\pm0.443\)    & \(0.587\pm0.105\) & 1.039                  \\
LkH$\alpha$ 310     & 0.0138  & 0.0002   & \(0.094\pm0.008\) & \(0.339\pm0.007\)  & \(0.251\pm0.018\)  & \(0.130\pm0.014\) & \(0.812\pm0.020\)  &                     & \(2.08\pm0.09\) &                  &  & \(6.246\pm0.690\)    & \(1.383\pm0.137\) & 0.677                  \\
HD 288313 A  & 0.1227  & 0.0018   &                   &                    &                    &                   &                    &                     &                 &                  &  &                      &                   &                        \\
PDS 277      & 0.0199  & 0.0003   &                   &                    &                    &                   &                    & \(3.550\pm0.319\)   &                 & \(10.24\pm0.09\) &  &                      &                   &                        \\
CR Cha       & 0.0686  & 0.0010   & \(0.343\pm0.006\) & \(1.142\pm0.013\)  & \(1.205\pm0.013\)  & \(0.819\pm0.011\) & \(1.359\pm0.019\)  & \(1.400\pm0.112\)   & \(1.75\pm0.02\) & \(7.85\pm0.08\)  &  & \(1.659\pm0.032\)    & \(2.388\pm0.022\) & 0.704                  \\
Ass ChaT2-21 & 0.1443  & 0.0022   & \(0.283\pm0.008\) & \(0.133\pm0.004\)  & \(0.134\pm0.004\)  & \(0.093\pm0.004\) & \(0.072\pm0.005\)  &                     & \(0.73\pm0.03\) &                  &  & \(0.774\pm0.063\)    & \(0.329\pm0.051\) & 1.251                  \\
DI Cha       & 0.0949  & 0.0014   & \(1.553\pm0.015\) & \(2.214\pm0.014\)  & \(2.260\pm0.023\)  & \(1.829\pm0.020\) &                    &                     & \(3.03\pm0.01\) &                  &  &                      & \(1.178\pm0.015\) & 0.957                  \\
CV Cha       & 0.0605  & 0.0009   & \(0.786\pm0.007\) & \(2.575\pm0.009\)  & \(2.082\pm0.012\)  & \(1.106\pm0.006\) &                    & \(2.840\pm0.199\)   & \(2.78\pm0.01\) & \(8.75\pm0.07\)  &  &                      & \(1.407\pm0.010\) & 0.742                  \\
Ass ChaT2-54 & 0.0521  & 0.0008   & \(0.067\pm0.002\) & \(0.035\pm0.001\)  & \(0.052\pm0.002\)  & \(0.028\pm0.001\) & \(0.155\pm0.008\)  & \(0.794\pm0.064\)   & \(0.27\pm0.03\) & \(7.52\pm0.08\)  &  & \(5.536\pm0.347\)    & \(0.418\pm0.047\) & 1.251                  \\
HD 135344B   & 0.1080  & 0.0016   & \(2.210\pm0.040\) &                    & \(1.177\pm0.002\)  & \(0.763\pm0.002\) & \(7.529\pm0.044\)  & \(25.600\pm3.072\)  & \(3.28\pm0.02\) & \(10.52\pm0.12\) &  & \(9.868\pm0.061\)    & \(0.345\pm0.018\) &                        \\
HT Lup  & 0.1583  & 0.0024   & \(1.295\pm0.006\) & \(2.533\pm0.054\)  & \(2.584\pm0.055\)  & \(2.047\pm0.034\) & \(3.448\pm0.015\)  & \(7.920\pm0.790\)   & \(2.28\pm0.01\) & \(8.81\pm0.10\)  &  & \(1.684\pm0.029\)    & \(1.581\pm0.017\) & 0.916                  \\
HD 142666    & 0.0875  & 0.0013   & \(2.429\pm0.004\) & \(4.780\pm0.006\)  & \(4.816\pm0.010\)  & \(3.508\pm0.012\) & \(5.508\pm0.036\)  & \(7.230\pm0.578\)   & \(3.61\pm0.00\) & \(9.33\pm0.08\)  &  & \(1.570\pm0.012\)    & \(1.444\pm0.004\) & 0.934                  \\
HD 142527    & 0.2063  & 0.0030   & \(6.526\pm0.019\) & \(11.447\pm0.033\) & \(11.663\pm0.017\) & \(6.227\pm0.010\) & \(31.518\pm0.039\) & \(10.500\pm1.260\)  & \(3.75\pm0.00\) & \(8.86\pm0.12\)  &  & \(5.062\pm0.010\)    & \(0.954\pm0.003\) & 1.037                  \\
HD 144432    & 0.0904  & 0.0013   & \(2.290\pm0.006\) & \(11.138\pm0.017\) & \(8.415\pm0.007\)  & \(4.088\pm0.013\) & \(7.573\pm0.049\)  & \(5.760\pm0.518\)   & \(3.51\pm0.00\) & \(9.12\pm0.09\)  &  & \(1.852\pm0.013\)    & \(1.785\pm0.004\) & 0.654                  \\
Haro 1-6     & 0.2157  & 0.0036   & \(0.593\pm0.011\) & \(0.263\pm0.020\)  & \(0.697\pm0.030\)  & \(0.415\pm0.015\) & \(4.157\pm0.080\)  &                     & \(1.10\pm0.02\) &                  &  & \(10.017\pm0.410\)   & \(0.700\pm0.041\) & 2.695                  \\
EM*SR 21     & 0.1098  & 0.0018   & \(1.029\pm0.014\) & \(1.569\pm0.010\)  & \(2.549\pm0.018\)  & \(2.306\pm0.022\) & \(25.425\pm0.343\) & \(33.800\pm3.718\)  & \(2.43\pm0.01\) & \(10.67\pm0.11\) &  & \(11.026\pm0.182\)   & \(2.241\pm0.017\) & 1.285                  \\
AK Sco       & 0.1045  & 0.0015   & \(1.148\pm0.007\) & \(3.760\pm0.032\)  & \(2.533\pm0.016\)  & \(1.269\pm0.010\) & \(4.237\pm0.037\)  & \(6.110\pm0.550\)   & \(2.60\pm0.01\) & \(9.01\pm0.09\)  &  & \(3.339\pm0.039\)    & \(1.105\pm0.010\) & 0.650                  \\
PDS 156      & 0.0519  & 0.0008   &                   &                    &                    &                   &                    & \(2.900\pm0.232\)   &                 & \(8.93\pm0.08\)  &  &                      &                   &                        \\
DI Cep       & 0.0250  & 0.0004   & \(0.409\pm0.007\) & \(1.485\pm0.027\)  & \(1.172\pm0.024\)  & \(0.710\pm0.020\) & \(1.675\pm0.021\)  &                     & \(3.04\pm0.02\) &                  &  & \(2.359\pm0.073\)    & \(1.736\pm0.033\) & 0.676                  \\
V395 Cep     & 0.0584  & 0.0009   & \(0.412\pm0.008\) & \(0.677\pm0.017\)  & \(0.681\pm0.017\)  & \(0.488\pm0.018\) & \(1.105\pm0.015\)  & \(1.550\pm0.155\)   & \(2.12\pm0.02\) & \(8.14\pm0.10\)  &  & \(2.264\pm0.089\)    & \(1.184\pm0.042\) & 0.950                  \\
\hline
\end{tabular}
\end{sidewaystable*}

\begin{table}
\caption{Index of PAH detections in the Spitzer spectra. The 'x' indicates where a detectable feature is identified.}
\label{tbl:PAH_detect}
\begin{tabular}{llllll}
\\
\hline \hline
Name & \multicolumn{5}{c}{PAH emission lines detections}  \\
  & $6.2\mu$m & $7.7\mu$m & $8.6\mu$m & $11.3\mu$m & $12.7\mu$m\\
\hline
BX Ari       &   &   &   &   &   \\
HBC 338      &   &   &   & x & x \\
LkH$\alpha$ 330     & x &   &   & x &   \\
RY Tau       &   &   &   &   &   \\
T Tau        &   &   &   &   &   \\
UX Tau A     &   &   &   & x & x \\
HQ Tau       &   &   &   &   &   \\
HBC 415      &   &   &   &   &   \\
SU Aur       &   &   &   &   &   \\
HD 34700     &   &   &   &   &   \\
CO Ori       &   &   &   &   &   \\
HD 35929     &   &   &   &   &   \\
PDS 115      &   &   &   &   &   \\
GW Ori       &   &   &   &   &   \\
V1650 Ori    & x &   &   &   &   \\
GX Ori       & x &   & x & x & x \\
RY Ori       &   &   &   &   &   \\
HBC 442      & x &   &   & x &   \\
SW Ori       &   &   &   &   &   \\
V1044 Ori    & x &   &   &   &   \\
EZ Ori       &   &   &   &   &   \\
Brun 252     &   &   &   &   &   \\
V2149 Ori    & x & x & x & x & x \\
Brun 555     &   &   &   &   &   \\
Brun 656     &   &   &   &   &   \\
V815 Ori     & x & x & x & x &   \\
PR Ori       &   &   &   &   &   \\
HD 294260    & x &   &   & x &   \\
BE Ori       &   &   &   &   &   \\
HBC 502      &   &   &   &   &   \\
LkH$\alpha$ 310     &   &   &   &   &   \\
HD 288313 A  &   &   &   &   &   \\
PDS 277      &   &   &   &   &   \\
CR Cha       &   &   &   &   &   \\
Ass ChaT2-21 &   &   &   & x &   \\
DI Cha       &   &   &   &   &   \\
CV Cha       &   &   &   &   &   \\
Ass ChaT2-54 &   &   &   & x &   \\
HD 135344B   &   &   &   & x &   \\
HT Lup       &   &   &   &   &   \\
HD 142666    & x & x & x & x &   \\
HD 142527    & x &   &   &   &   \\
HD 144432    & x &   &   &   &   \\
Haro 1-6     & x & x &   & x & x \\
EM*SR 21     & x &   & x & x &   \\
AK Sco       & x &   &   &   &   \\
PDS 156      &   &   &   &   &   \\
DI Cep       &   &   &   &   &   \\
V395 Cep     &   &   &   &   &   \\
\hline
\end{tabular}
\end{table}


%
%

\bibliographystyle{aa}
\bibliography{Paper1}

\begin{appendix} 

\section{High RUWE in Final sample}

Here we describe the distance determination for those sources where the \textit{Gaia} RUWE is larger than 1.4 or where no \textit{Gaia} parallax was available.

\begin{itemize}
    \item
     For LkH$\alpha$ 330 we confirm the parallax by using the early \textit{Gaia} DR3 release and compare the proper motion with stars in a 10 arc minute radius around the star with similar parallaxes. LkH$\alpha$ 330, located in Perseus, has a distance of 308 pc \citep{2018AJ....156...58B}. Examining the DR3 proper motion puts it in a group of 34 stars. The average proper motion of this group is RA 4.19 mas yr$^{-1}$ ($\sigma =0.37$) and DEC -6.00 mas yr$^{-1}$ ($\sigma = 0.61$), compared to the proper motion of LkH$\alpha$ 330, RA 4.58 mas yr$^{-1}$ and DEC -5.66 mas yr$^{-1}$ respectively. Within the group the range of parallaxes are 2.96 to 3.99 mas. The DR2 parallax of LkH$\alpha$ 330 is 3.22. We find that this strengthens the estimated distance from \citet{2018AJ....156...58B}.
    \item
     Using the same approach for HQ Tau we find a similar proper motion, RA 10.87 mas yr$^{-1}$ DEC -18.98 mas yr$^{-1}$, as a small group of 7 stars with an average proper motion of RA 10.68 mas yr$^{-1}$ ($\sigma = 0.5$) and DEC -17.55 mas yr$^{-1}$ ($\sigma = 1.38$). The spread of parallaxes of these stars range from 6.09 to 6.71 with HQ Tau parallax at 6.2. We see it as likely that these stars are a small co-moving group. This is also strengthened by that the \textit{Gaia} DR2 parallax lies within 1-{$\sigma$} of the parallax determined from very-long-baseline-interferometry (VLBI) which associates HQ Tau with the HP Tau/G2 group \citep{2020ApJ...889..175R}. HQ Tau association with this group further strengthens the parallax and we therefore use the distance according to \citet{2018AJ....156...58B} despite of the large RUWE.
    \item
     The distance for RY Tau differs greatly between \textit{Gaia} DR1 and DR2 measurements  (corresponding to 176 pc and 442pc respectively). The previous \textit{Hipparcos} measurement puts RY Tau at a distance of 133 pc. The inverted EDR3 parallax suggests that the distance is 138 pc but the RUWE is still very high ($\sim $13). \citet{2019AA...628A..68G} argues using the proper motion of RY Tau that it most likely being a member of the Taurus star forming region and that a distance based on the inverted \textit{Hipparcos parallax} of 133 pc is likely correct. We therefore use the \textit{Hipparcos} distance for RY Tau.
    \item
     Due to the similarity between the \textit{Gaia} DR2 and \textit{Hipparcos} parallax for HT Lup, 6.48 and 6.29 mas respectively, we decide to use the value from \citet{2018AJ....156...58B}. 
    \item
     GW Ori lies in the $\lambda$ Ori association and comparing the distance of the association, $403^{+13}_{-8}$ pc with that of \citet{2018AJ....156...58B}, $398^{+10.6}_{-10.1}$ pc, strengthens the \textit{Gaia} measurement despite the high RUWE. We have decided to use the \textit{Gaia} based distance.  
    \item
     EZ Ori was identified by radial velocity as a member of ONC-23 group of stars ($399^{+26.5}_{-23.4}$ pc) \citep{2018AJ....156...84K}. This confirms the distance from \citet{2018AJ....156...58B} ($361^{+40.7}_{-33.3}$) pc.
    \item
     The stars Ass Cha-T2 21 and Ass Cha-T2 54 are members of the Chameleon I molecular could. The estimated distances from the \textit{Gaia} parallax by \citet{2018AJ....156...58B}, $164.77^{+3.95}_{-3.77}$ and $202.95^{+19.19}_{-16.18}$ respectively, both overlap with the determined distance to Chameleon I $179^{+11}_{-10}$  \citep{2018A&A...610A..64V}. We therefore use the distances in \citet{2018AJ....156...58B} for both stars.
    \item
     PR Ori is a member of the Lynds 1641 molecular cloud \citep{1999MNRAS.310..331M}. A distance determination to Lynds 1641 was made by \citet{2019AA...624A...6Y} using \textit{Gaia DR2} parallaxes. The cloud distance $408_{-4}^{+4}$ pc is slightly larger than that derived by \citet{2018AJ....156...58B} ($\sim 374$ pc). We choose to adapt the cloud distance rather then the \textit{Gaia} derived distance as the distance to PR Ori.
    \item
     The parallax of T Tau was measured by \citet{2018ApJ...859...33G} using the \textit{VLBI} indicating a distance of 148.7($\pm 1.0$) pc which corresponds well to the \textit{Gaia} based distance of 143.74$_{-1.21}^{+1.22}$ pc.
    \item
     For 3 stars no \textit{Gaia} DR2 parallaxes were available and we searched the literature for distances. V2149 Ori are a member of the Orion Nebula Complex and we used the distance 388$\pm{5}$ pc \citep{2019AA...622A..72V}. For UX Tau we used the distance given in \citet{2019ApJ...872..158A}, 139.4$\pm{1.96}$ pc. And finally for HD 288313A, located in Lynds 1622 at a distance of 418$\pm{17}$ \citep{2020AA...633A..51Z}.
    
\end{itemize}

\section{Model SEDs}

A key difference between the classical HAeBe stars and the precursors we are identifying in this paper it the temperature of the start itself. In order to check what difference in the measured disk quantities is down to only this change, and how this compares to the effects of a variation in the scale height of the inner disk, we are computing a simple grid of SEDs. For this purpose, we consider a 2 $M_\odot$ PMS star using a 
calculated model SEDs for a passively irradiated disk 
using stellar parameters taken from the \cite{2000AA...358..593S} evolutionary tracks (see table~\ref{tab:A1}). We use the temperature range between $T_{\rm eff}$=4900K and $T_{\rm eff}$=9000K. We use the Monte Carlo radiative transfer code MCMax \citep{2009AA...497..155M}, and the DIANA model set-up \citep{2019PASP..131f4301W} to calculate the SEDs (table~\ref{tab:A2}). We fix the inner rim dust temperature to 1500~K, which means that the disk inner radius increases with increasing stellar luminosity as the star approaches the Zero Age Main Sequence. We ignore the possible effects of an evolving dust grain population, such as grain settling that may vary in time, inward drift, gaps, holes and planet formation on the dust distribution. We chose two values for the disk scale height at 0.4 AU (0.041 and 0.08 AU) to study the effect of the inner disk scale-height on the resulting SED.  We derive model values for the [$F_{30}/F_{13.5}$] flux ratio and the 7 $\mu$m excess. 

\begin{table}
\caption{Stellar parameters following the evolutionary track of a 2 M$_{\odot}$ PMS star \citep{2000AA...358..593S}}\label{tab:A1}
\centering
\begin{tabular}{lcccc}
\hline \hline
$T_{\rm eff}$ & $L_*$ & $R_*$ & log g & $R_{inner}$  \\
K & $L_{\odot}$ & $R_{\odot}$ & cm/s$^2$ & AU \\
\hline
4910 & 6.90   & 3.44 & 3.67 & 0.184 \\
5560 & 7.94   & 2.89 & 3.82 & 0.197 \\
6280 & 16.50  & 3.26 & 3.71 & 0.284 \\
7275 & 22.40  & 2.85 & 3.83 & 0.331 \\ 
9080 & 18.49  & 1.68 & 4.29 & 0.301 \\
\hline
\end{tabular}
\end{table}

\begin{table}
\caption{Disk parameters used for the evolution model SED.}\label{tab:A2}
\centering
\begin{tabular}{lc}
\hline \hline
parameter & value (unit) \\
\hline
$R_{inner}$ & see table A.1 \\
$R_{tap}$ & 100~AU  \\          
$R_{out}$ & 475~AU  \\           
$M_{dust}$ & 2.18E-04~$M_{\odot}$ \\ 
gas2dust & 100    \\
$\epsilon$ & 0.66      \\     
$\gamma_{exp}$ & 1.34    \\     
$\beta$ & 1.095       \\     
sh & 0.041~AU           \\     %
$R_{sh}$ & 0.4~AU          \\     
amin & 5.0E-02~$\mu$m         \\     
amax & 3.0E+03~$\mu$m         \\     
apow & 3.50         \\     
$\alpha_{settle}$ & 0.01   \\     
\hline
\end{tabular}
\end{table}

\begin{figure}
\label{fig:modelSEDs}
\centering
\includegraphics[width=8.8cm,clip]{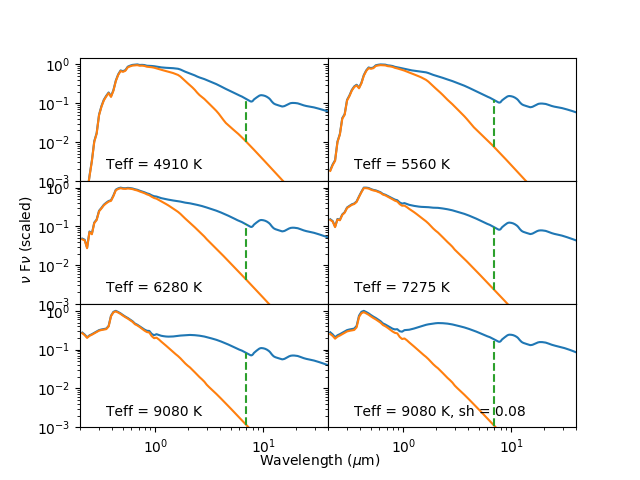}
\caption{SEDs of a 2 $M_{\odot}$ PMS star surrounded by a passively heated gas/dust disk. We use a range of luminosities and temperatures (see table~\ref{tab:A1}) representative of the evolution along a \cite{2000AA...358..593S} evolutionary track. The vertical flux scale is normalized to its maximum value for each model. Disk parameters are listed in table \ref{tab:A2}. The dashed line indicates the 7 $\mu$m excess point. The lower right panel shows the effect of increasing the disk scale height at 0.4 AU from 0.041 to 0.08 AU, for the $T_{\rm eff}$ = 9080~K model.}
\label{appfig}
\end{figure}

Figure \ref{fig:modelSEDs} shows our small model grid, for which  7 $\mu$m excess values increase from 2.7 to 4.6 magnitudes as $T_{\rm eff}$ increases from 4910 to 9080~K. Increasing the 0.4~AU disk scale height from 0.041 to 0.08~AU changes these numbers to 3.6 and 5.6 magnitudes, respectively. The [$F_{30}/F_{13.5}$] remains relatively constant at values between 1.7 and 2.1 for all models. We conclude that an increase in $T_{\rm eff}$ results in a blueward shift of the stellar SED and therefore an increase in 7 $\mu$m excess of about 2 magnitudes. There seems no need for changes in the inner disk structure to explain the difference in the maximum 7 $\mu$m excess between IMTT and HAeBe stars. At the same time, these (gapless) models, as expected, do not account for the wide range in observed [$F_{30}/F_{13.5}$]. We refer to \cite{2009AA...502L..17A} for a full discussion on the effect of model parameters on the [$F_{30}/F_{13.5}$] flux ratio. 

\section{Individual sources}
    In this section we discuss the gas rich disk of stars in the sample which we have found spatially resolved data in scattered light or at millimeter wavelengths in published literature. For each star we check if the spatially resolved data support the Group I or Group II classification based on the SED (section 5). In scattered light and at millimeter wavelengths the Group I disks can be identified having large disk cavity ($\geq 5AU$) while Group II disks are smaller continuous disks lacking large cavities \citep{2017AA...603A..21G}.

\subsection{LkH${\alpha}$ 330} 
    The first detection of the disk was made in IRAS survey data \citep{1990ApJS...74..575W,1992ApJS...78..239W}. The disk has an inner cavity first detected by \citet{2007ApJ...664L.107B}. Using sub-millimeter imaging the cavity was determined to be $\sim$50-70 AU \citep{2009ApJ...704..496B,2011ApJ...732...42A} with the outer disk extending to $\sim$125AU \citep{2009ApJ...704..496B}. This is consistent with our SED classification for the disk as a Group I source. \citet{2013ApJ...775...30I}, using millimeter-wave interferometry found a lopsided ring of $\sim$100 AU in the 1.3mm dust continuum suggesting the cause to be planet-disk interaction. Observations with the SUBARU in H-band and at 0.87mm revealed a spiral structure in the disk \citep{2016AJ....152..222A} explained by the possibility of an unseen planet companion in the disk. Revisiting the star with SUBARU \citet{2018AJ....156...63U} found the gap in scattered light to be $\sim$54 AU also clearly detecting the spiral arms.

\subsection{RY Tau}
    An active X-ray source \citep{2016ApJ...826...84S} with a visible jet \citep{2009AA...493.1029A,2020AA...633A..82G} RY Tau has a spectral slope of a Group II disk. ALMA observations reveal a 60 AU disk with a small central cavity two rings (at $\sim$18 AU and $\sim$50 AU) a gap ($\sim$ 43 AU) \citep{2018ApJ...859...32P,2018ApJ...869...17L} while \citet{2020ApJ...892..111F} report an inner cavity of 27 AU.In H-band observations with SUBARU a disk similar in size to the ALMA observations are detected but the cavity, rings and gap are not detected \citep{2013ApJ...772..145T}.

\subsection{UX Tau} 
    UX Tau is a triple system with components A, B and C. The B component also is a binary. The disk around UX Tau A based on its SED is a Group Ia disk. Already by \citet{2012PASJ...64..124T} classified as pre-transitional. The disk is strongly polarized, geometrically thin and extends to 120AU \citep{2012PASJ...64..124T}. An inner cavity was detected by \citet{2007ApJ...670L.135E} and later determined to a size of 25-30AU \citep{2011ApJ...732...42A,2012PASJ...64..124T,2014AA...564A..51P} consistent with our classification. The disk has a spiral structure visible both in scattered light and in gas continuum \citep{2020ApJ...896..132Z,2020AA...639L...1M}. The spiral arms likely comes from tidal interaction with UX Tau C towards which one of the arms also extend \citep{2020ApJ...896..132Z}. Zapata et al. further also detects a disk around UX Tau C, material they suggest is probably coming from the interaction with the disk of the A component.

\subsection{HQ Tau} 
    The dusk around HQ Tau was first classified as a possible debris disk \citep{2005ApJ...631.1134A} later reclassified as a transitional disk \citep{2009ApJ...703.1964F, 2010ApJS..186..111L}. ALMA observations of the disk \citep{2019ApJ...882...49L} reveals a compact disk with indications of dust depletion towards the inner disk. Our classification as a Group IIa disk fit well with this observation. 

\subsection{SU Aur} 
    \citet{2005ApJ...629..881H} using IRAC was fist to detect the infrared excess of SU Aur. \citet{2005ApJ...622..440A} determined the inner diameter of the dusty disk to be $\sim$0.2 AU based modeling using the observations of 2MASS J, H and K band. The disk was studied with both ALMA \citep{2015ApJ...806L..10D} and SUBARU \citep{2017AJ....153..106U, 2019AJ....157..165A} which shows the disk to be disturbed and having a long tail-like structure extending a thousand AU possibly caused by the interaction of a flyby by a sub stellar companion or the ejection of a small object from the disk. \citet{2019AA...627A..36L} using the CHARA array performed interferometric observations of SU Aur . They found the disk starting at 0.15 AU extending out to 100 AU with an inclination of 50$^{\circ}$. Based on the SED we find this disk to be a Group I source.

\subsection{HD 34700} 
    HD 34700 is a multiple system where the A component a intermediate binary system, $2 M_{\odot}+2 M_{\odot}$ with a separation of 0.69 AU. \citep{2003AA...408L..29A,2005AA...434..671S,2019ApJ...872..122M}. The separation of the B an C components to the A component is 5.18" and 9.17" respectively \citep{2005AA...434..671S}. HD 34700A was imaged by the Gemini Planet Imager, GPI, \citep{2019ApJ...872..122M} which showed a very prominent transition disk inclined at 41$^{\circ}$, with a series of spiral arms, surrounding the binary  \citep{2019ApJ...872..122M}. This is consistent with our classification from the SED as a Group I disk. The disk as a large cavity and begins at 175 AU from the center of the system and extends out to 500 AU. The cavity is too large to be a consequence of only the two stellar components and could be the effect of a massive planet companion \citep{2019ApJ...872..122M}.

\subsection{GW Ori} 
    GW Ori s a triple system \citep{2011AA...529L...1B} consisting of a spectroscopic binary (GW Ori AB) with a separation of $\sim$1 AU \citep{1994ARAA..32..465M} and a C component, separated $\sim$8AU \citep{2011AA...529L...1B}. The disk is a circumtriple disk with a dust component extending to $\sim$400 AU and a gas component extending to $\sim$1400 AU. The disk is gapped at 25-55 AU \citep{2014AA...570A.118F} and spatially resolved ALMA observations shows it to have three dust rings at 46, 188 and 338 AU \citep{2020ApJ...895L..18B}, the latter the largest dust ring presently known in a protoplanetary disk. Our Group I SED classification fit well with this data. The CO kinematics of the disks suggest that there is a miss-alignment of the inner disks spin axis in respect to the outer disk plane \citep{2020ApJ...895L..18B}.  

\subsection{V2149 Ori} 
    V2149 is a known binary star \citep{2006AA...458..461K}, G0+F7 \citep{2012AA...540A..46D}. \citet{2013ApJ...769..149K} classified the disk, based on that the SED was indicative of a central cavity, as transitional and estimated the radius of the disk to 138AU. We find V2149 Ori to be a Group II disk but the uncertainty of the spectral index puts it in the border between Group I and Group II disks. Recent scattered light observations with the SPHERE instrument (Valegard et al. paper in preparation) do not detect any disk. This either means the disk is small or self-shadowed suggesting the Group II classification to be correct.

\subsection{CR Cha} 
     The first detection of a disk around CR Cha was made by \citet{1993AA...276..129H} that classified the disk as Lada Class II. \citet{2011ApJ...728...49E} found that the IRS spectra best fitted a pre-transitional disk model which is consistent with our classification. A cavity in the disk was suggested by \citet{2014AA...564A..51P} using SED modeling. \citet{2018AA...617A..83V} used interferometric fitting from MIDI observations to derive an inner disk radius of $\sim$1.3 AU. Observations by \citet{2020ApJ...888...72K} with ALMA Band 6 show no cavity, instead a gap in the dust continuum at $\sim$90 AU with a width of $\sim$8 AU and a dust ring at $\sim$120 AU. The SED of CR Cha has a  spectral slope that places it as a Group IIa source witch is consistent with this continuous inner disk.

\subsection{DI Cha} 
    DI~Cha is a quadruple system with two binaries, one with a G2 and M6 star (the latter possibly a brown dwarf), and the second binary consisting of two M5.5 dwarfs \citep{2013AA...557A..80S}. The angular separation between the two sets of binary stars is 4.6 arcsec. \cite{2020ApJ...895..126H} find an upper limit of 0.12 arcsec from ALMA continuum imaging (90 per cent light radius) for the mm continuum emission centered on the G star, and \cite{2015AA...581A.107M} derive a size of 14.1 milli-arcsec from N-band interferometry. The ALMA and MIDI data taken together suggest a compact disk. \cite{2018AA...617A..83V} from N-band interferometry find that the inner disk size is not compatible with a continuous disk, but may contain a disk gap on spatial scales of AU.  The available imaging is consistent with a classification as a GII source.

\subsection{CV Cha}  
    CV Cha is a visual binary with an M1 companion at a distance of 11.4 arcsec \citep{1993AA...278...81R}. \cite{2020ApJ...895..126H} resolve the disk in mm dust continuum with ALMA and find a size of 0.14 arcsec (90 per cent light radius). \cite{2015AA...581A.107M} resolve the inner disk in the N band and find a half light radius of 6.1 milli-arcsec.  The photosphere of CV Cha is heavily veiled, with the accretion shock covering 20-40\% of the stellar surface \citep{2014ApJ...786...97H}. This may introduce some uncertainty in the derived stellar parameters. The available spatial information is consistent with a classification as a GII source.

\subsection{HD 135344B} 
    HD135344B is a visual binary with the A0 star HD135344A (angular separation 21 arc seconds). Many studies are devoted to the geometry of its circumstellar disk. \cite{2011ApJ...732...42A} spatially resolve the disk at mm wavelengths using the SMA and find a dust cavity with a radius of 46 AU. \cite{2012ApJ...748L..22M} resolve the disk in scattered light and found two spiral arms at 70~AU from the star. \cite{2013AA...560A.105G} used NACO at the VLT to measure the radius of the dust in scattered light (probing small grains in the disk surface) and found a radius of 28~AU, much smaller than the 46~AU found for the large mid-plane grains. \cite{2016AA...595A.113S} and \cite{2017ApJ...849..143S} find time variable shadows cast on the outer disk, probably resulting from changes in the inner disk structure.  \cite{2016ApJ...832..178V} resolved the mm dust into a ring at 50~AU and an asymmetric structure at 70~AU (see also \cite{2018AA...619A.161C}). These observations convincingly show that HD135344B is a transitional disk (GI), in which planet formation is the likely cause of the observed disk geometry. 

\subsection{HT Lup} 
    HT Lup is a triple system, HT Lup A, B and C. We classify the disk as a Group II which is consistent with DSHARP ALMA continuum images that show a small disk with spiral arm structure around the A component that spans roughly 30AU \citep{2018ApJ...869L..41A,2018ApJ...869L..44K}. The B and C component have disks of 5 and 9 AU respectively \citep{2018ApJ...869L..44K}. The disk has also been observed by SPHERE \citep{2020AA...633A..82G} and the scattered light signal is consistent with the disk being observed by ALMA for HT Lup A. The disks around the B and C components are not resolved in the scattered light images, neither is the spiral arm structure around the primary component.

\subsection{HD 142666} 
    DSHARP ALMA continuum images of HD~142666 reveal the presence of relatively narrow rings at distances between 6 and 40 AU, and an outer radius of about 90 AU  \citep{2018ApJ...869L..41A,2018ApJ...869L..42H}. The innermost 6 AU seem devoid of large, cold grains \citep{2018ApJ...860....7R}. In scattered light the disk is detected although weakly \citep{2017AA...603A..21G}. This star was classified as group II by \cite{2001AA...365..476M}. \cite{2009AA...502..367S} find evidence from near-IR interferometry for a gap at 0.5 AU. A comparison of disk size in the near- and mid-infrared by \cite{2019AA...632A..53G} strongly suggests that there is substantial inner disk structure. The classification of HD~142666 as a group II source can still be understood by the fact that the gap dimension is too small to cause a significant imprint on the SED, which was the basis for the classification. The case of HD~142666 illustrates that the group~I and group~II classification by  \cite{2001AA...365..476M}, while providing a useful separation between "self-shadowed" disks and (pre-)transitional disks with gaps on scales of 10 AU or more, fails to identify inner disk gaps on scales of AU.  Nevertheless, the outer disk does not receive much direct stellar light, as evidenced by the weak signal detected with SPHERE \citep{2017AA...603A..21G}. 

\subsection{HD 142527} 
    HD 142527 is perhaps the prototypical GI disk source with a large ($\sim$130 AU) gap  \citep{2006ApJ...636L.153F,2006ApJ...644L.133F,2008Ap&SS.313..101O,2013Natur.493..191C,2014ApJ...781...87A,2014ApJ...791L..37R}. The prominent shadows detected in scattered light are explained by a highly inclined inner disk \citep{2015ApJ...798L..44M}. SMA and ALMA data reveal a horseshoe distribution of dust in the outer disk, usually interpreted as a dust trap \citep{2008Ap&SS.313..101O,2013Natur.493..191C}; see \citep{2013Sci...340.1199V}. A stellar mass companion was detected close to the outer radius of the inner disk \citep{2012ApJ...753L..38B,2016AA...590A..90L}, which itself is surrounded by an accretion disk. Our SED based classification is confirmed by these observations.

\subsection{Haro 1-6} 
    Haro 1-6 is a Group I disk based on its SED with one of the highest [F$_{30}$/F$_{13.5}$] in the sample. Images in scattered light taken by SPHERE only marginally detect a disk signal around the edges of the coronograph \citep{2020AA...633A..82G} with several bright filaments extending around the star. \citet{2020AA...633A..82G} suggest that these filaments does not have anything to do with the formation of Haro 1-6. No mm emission is detected around Haro 1-6 \citep{2019MNRAS.482..698C}. \citet{2008RMxAC..34...14L} suggests Haro 1-6 to be a spectroscopic binary. The PAH emission could have its source either from the strong far-UV and X-ray radiation coming from the binary excites the PAHs in a disk in its last stages of clearing or from a small scale photodissociation region \citep{2009ApJ...703..252J}.

\subsection{EM*SR 21}
    SR21, at a distance of \(137.86_{-1.07}^{+1.10} \)\,pc \citep{2018AA...616A...1G} has a binary companion in a wide orbit \citep{2003ApJ...591.1064B}.  It might also be a compact binary with the projected distance of about 0.1" \citep{2009ApJ...698L.169E}. SR21 has been classified as a transition disk already based on the shape of the spectrum seen by Spitzer \citep{2007ApJ...664L.107B}. \citet{2011ApJ...732...42A} and \citet{2014ApJ...783L..13P} showed the presence of a large cavity in submm continuum emission, again consistent with a transitional disk. Recently, \citet{2020AA...636L...4M} presented combined data from ALMA in Band 3 and SPHERE polarimetric images in the H band showing a large cavity, a bright ring peaking at 53 au, and spiral structure visible in the scattered light observation inside the main ring, making SR21 a rather unique object. They also show the presence of a kinked spiral connecting the inner and outer disk, matching hydrodynamical predictions of a planet carving the gap and pinpointing its likely position.

\subsection{AK Sco} 
    AK Sco is a double lined spectroscopic binary \citet{1989Msngr..55...45A} comprised of a F6+F6 couple with a separation of $\sim$0.16 AU \citep{2015AA...574A..41A}. The circumbinary disk around AK Sco was detected first time by \citet{1996ApJ...458..312J}. The overall shape of the SED points towards a Group II disk, but with its steep mid-infrared spectral slope, with $[F_{30}/F_{13}] \sim 3.3$, our criteria classify it as a Group Ia disk. It has a strong silicate feature (see figure \ref{fig:C-GIsil}) that requires either a different dust opacity or an extreme grain distribution. It is one of the smallest disks detected in polarized light \citep{2017AA...603A..21G}. The inner rim of the disk is located at $\sim$0.58 AU \citep{2015AA...574A..41A} The polarized light observations show the radius of the disk to be $\sim$13-40 AU \citep{2016ApJ...816L...1J} with sharp features either from an eccentric ring or a spiral arm structure and confirms the compact nature $\sim$14 AU in previous observations by ALMA \citep{2015ApJ...806..154C}.

\end{appendix}

\end{document}